\documentclass[prl,twocolumn,amsmath,amssymb,floatfix,footinbib,bibnotes]{revtex4-1}
\pdfoutput=1

\usepackage[colorlinks=true,citecolor=blue,linkcolor=blue]{hyperref}
\usepackage{amsmath}
\usepackage{graphicx}
\usepackage{dsfont}
\usepackage{changepage}
\usepackage{fancyhdr}
\usepackage{amsthm, amssymb}
\usepackage{floatflt}
\usepackage{float}
\usepackage{array}
\usepackage[usenames,dvipsnames]{color}

\newcommand{\be}{\begin{align}}
\newcommand{\ee}{\end{align}}

\def \ua{{\uparrow}}
\def \da{{\downarrow}}
\def \be{\begin{equation}}
\def \ee{\end{equation}}
\def \ba{\begin{array}}
\def \ea{\end{array}}
\def \bea{\begin{eqnarray}}
\def \eea{\end{eqnarray}}
\def \nn{\nonumber}
\def \e{{\epsilon}}

\def \a{{\alpha}}
\def \t{{\theta}}
\def \b{{\beta}}
\def \g{{\gamma}}
\def \D{{\Delta}}
\def \d{{\delta}}
\def \w{{\omega}}
\def \s{{\sigma}}

\def \ve{{\varepsilon}}

\def \G{{\Gamma}}
\def \z{{\zeta}}

\def \ba{\begin{align*}}
\def \ea{\end{align*}}

\newcounter{indice}

\def \mrm{\mathrm}

\def \bs{\boldsymbol}

\begin{document}

\title{Superconductivity at very low density: the case of strontium titanate   }

\author{Jonathan Ruhman and Patrick A. Lee \\
{\small \em Department of Physics, Massachusetts Institute of Technology, Cambridge, MA 02139 USA}}
\begin{abstract}
Doped strontium titanate becomes superconducting at a density as low as $n = 5\times 10^{17}\,\mrm{cm}^{-3}$, where the Fermi energy is orders of magnitude smaller than the longitudinal-optical-phonon frequencies. In this limit the only optical mode with a frequency which is smaller than the Fermi energy is the plasmon. In contrast to metals, the interaction strength is weak due to screening by the crystal, which allows the construction of a controllable theory of plasmon superconductivity. We show that plasma mediated pairing alone can account for the observed transition temperatures if the screening by the crystal is reduced in the slightly doped samples compared with the insulating ones. This mechanism can also explain the pairing in the two-dimensional superconducting states observed at surfaces and interfaces with other oxides. We also discuss unique features of the plasmon mechanism, which appear in the tunneling density of states above the gap.
\end{abstract}
\maketitle
\noindent
{\it Introduction --}
The BCS theory very successfully explains superconductivity in metals. The essential attraction between electrons, according to this theory, is generated by exchange of phonons, which have a characteristic frequency $\w_D$. A crucial condition for the applicability of the theory is the 'retardation' condition, namely that $\w_D \ll \e_F$, where $\e_F$ is the Fermi energy~\cite{bogoliubov1960new}. This condition holds in almost all known conventional superconductors and seems to be a universal property.

An outstanding exception is doped strontium titanate (SrTiO$_3$). Free charge carriers in this material are achieved by inducing oxygen vacancies or doping with elements such as La or Nb.
Superconductivity is typically observed at temperatures lower than a few hundreds of Milikelvins~\cite{Schooley1964}. The transition temperature exhibits a dome shape as a function of carrier concentration, which extends to surprisingly low densities~\cite{Koonce1967,Binnig1980,Binnig1981,Lin20133}.
Recently it has been reported that superconductivity extends to densities as low as $n_{3d} = 5 \times 10 ^{17}\,\mrm{cm^{-3}}$ where the Fermi energy is $\e_F \sim$1meV~\cite{Lin2014}.
In this situation $\e_F$ is certainly not greater than $\w_D$, and therefore BCS theory does not apply. The natural question is therefore, why is strontium titanate superconducting at such a low density?

SrTiO$_3$ also exhibits non-trivial phenomena in its insulating state. Upon cooling, the polarizability of this material diverges with a Cuire-Weiss behavior signaling a ferroelectric instability. However, this behavior is cutoff before the instability is reached and eventually strontium titanate remains paraelectric all the way down to zero temperature~\cite{weaver1959,Muller1979}. The soft {\it transverse} optical phonon associated with the instability leads to a huge dielectric constant which, for our purposes, can be approximated by a single resonance model
\be
\ve(i\w) = \ve_{\infty} + (\ve_{0}-\ve_{\infty}){\w_T ^2 \over \w_T^2 +\w^2}\label{e(w)}
\ee
where $\w_{T} \approx 1.9$ meV is the frequency of the transverse mode at $T = 0$~\cite{Vogt1995,Vogt1981}, $\e_\infty = 5.1$~\cite{Kamaras1995} and $\e_0 \approx 2\times 10^4$~\cite{weaver1959}.
The Coulomb interaction $V(\w,r) =  e^2/\ve(\w) r$ has a pole at the frequency of the {\it longitudinal} phonon mode, $\w_L$, which is related to the transverse one by the Lyddane-Sachs-Teller (LST) relation $\w_L = \sqrt{\ve_0/\ve_\infty}\,\w_T$.
\begin{figure}
\centering
\includegraphics[width=8. cm,height=8.cm]{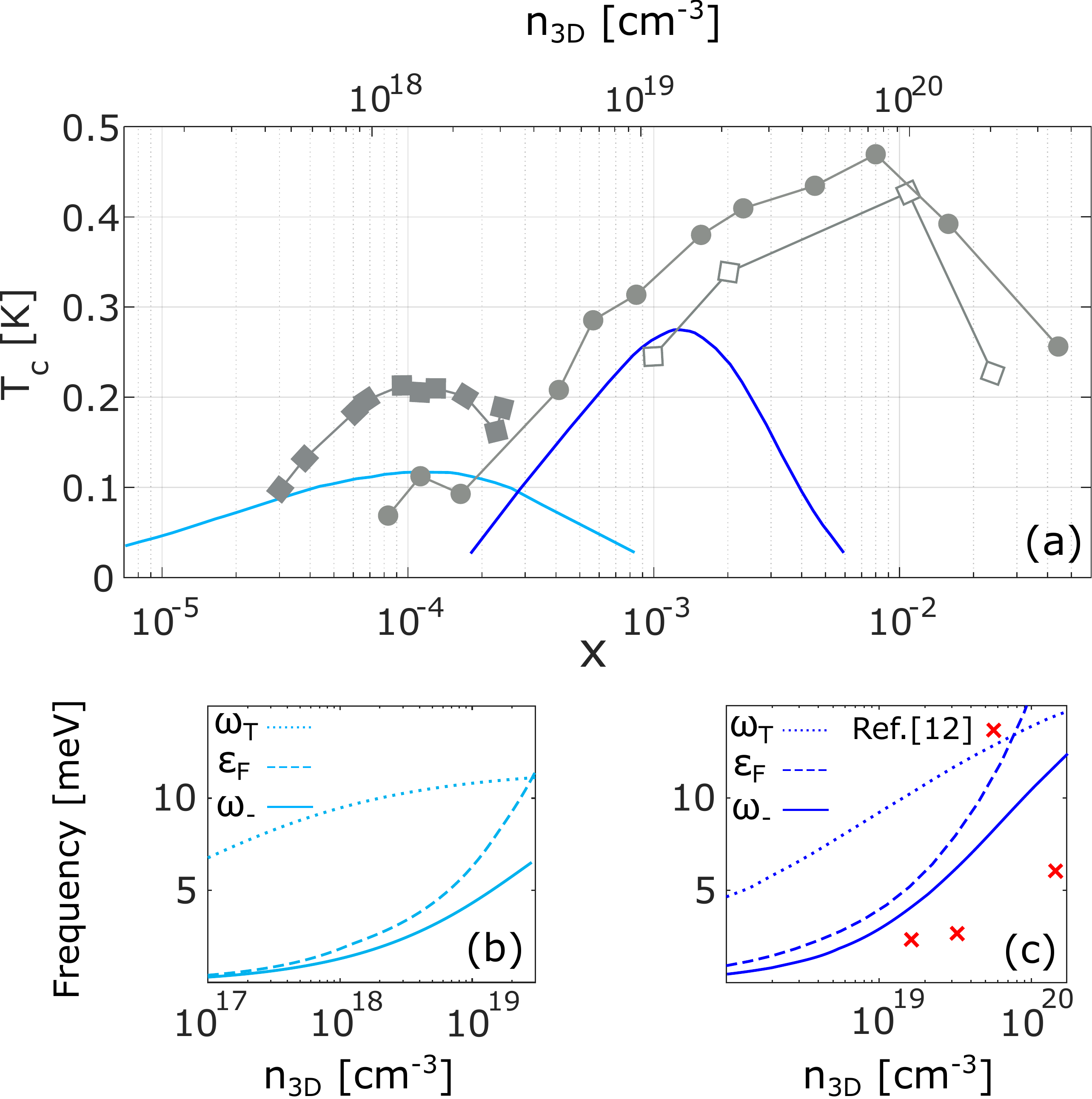}
\caption{(a) The transition temperature vs. electron doping. The blue (cyan) curve corresponds to $\eta = 0.3$, $m_1 = 4 m_e$ ($\eta = 0.4$, $m_1 = 2 m_e$). The soft transverse optical mode $\w_T$ is chosen as a parameter and is shown together with $\w_0$ and $\e_F$ in (b) for the cyan $T_c$ curve and in (c) for the blue curve. The $\mathds{x}$'s in red mark the measured $\w_T$ values~\cite{vanderMarel2008}.  The grey curves are experimental data points, the squares are from Ref.~\cite{Lin20133} (filled - oxygen reduced and empty - Nb doped) and the circles are from Ref.~\cite{Schooley1964}.   }\label{fig:domes}
\end{figure}

Gurevich, Larkin and Firsov (GLF) \cite{Gurevich1962} were the first to point out the potential importance of the longitudinal phonon mode to superconductivity.
They considered the attractive electron-electron interaction mediated by long-range Coulomb potentials induced this mode.  Therefore, in their theory the frequency $\w_L$ plays the role of $\w_D$ in the BCS theory.
For the parameters used in Eq.~\ref{e(w)} one obtains $\w_L = \sqrt{\ve_0/\ve_\infty}\,\w_T \approx 100\,\mrm{meV}$, such that $\w_L \gg \e_F$. It is therefore not possible to use the longitudinal mode to explain the superconducting state of SrTiO$_3$.

Early theoretical studies of superconductivity in SrTiO$_3$~\cite{Koonce1967} assumed multiple valleys and emphasized the importance of intervalley phonon scattering. These assumptions are now known to be incorrect.  Later, Takada~\cite{Takada1980}, added dynamical electronic screening to the GLF model~\cite{Gurevich1962} and used the theory of  Ref.~\cite{kirzhnits1973description} to calculate $T_c$, again with very good agreement with experiment.
Interestingly, Takada  proposed that plasma oscillations participate in mediating the attractive interactions.
However, his theory is uncontrolled because he incorporates the longitudinal phonon and the plasmon as mediators of a attractive interaction even when their frequencies are significantly larger than the Fermi energy, which is known to be problematic~\cite{Grabowski1984,Grimaldi1995}.
Indeed, his attraction is mainly generated by the higher frequency mode, i.e. the phonon at low density and the plasmon at higher density (see for example the conclusions in Ref.\cite{Klimin2014}).

We would also like to note two  recent studies of phonon-mediated superconductivity in SrTiO$_3$. Ref.~\cite{gor2015phonon} argued that multiplicity of longitudinal optical phonons leads to instantaneous attraction between electrons.
In the SI we show that this is not possible in the standard picture of screening due to polar phonons.
Ref.~\cite{Edge2015} tied the dome shape of $T_c$ to softening of the ferroelectric mode observed in DFT calculations.
But the coupling to the transverse mode is too weak when the density of states is so small.

In this paper we revisit the question of superconductivity in SrTiO$_3$ in light of new data using the Eliashberg theory. Our approach is to construct a controllable theory and focus on the extreme low density limit where the open questions are clearest.
In this limit the screened plasma frequency is the only resonance of the interaction that occurs below the Fermi energy, and therefore we agree with Ref.~\cite{Takada1980} that it is an important mechanism for pairing of electrons.
Our theory is controlled by weak coupling to the plasmon, which is provided by the screening of the crystal.
However, unlike Ref.~\cite{Takada1980}, we find that the coupling is too weak if the dielectric constant measured in insulating SrTiO$_3$ $\ve \simeq 2\times 10^{4}$~\cite{weaver1959,Vogt1981,Kamaras1995} is used.
Given our belief that the plasmon is the only low lying mode that is capable of inducing pairing, we find that the only way to obtain a realistic transition temperature at the lowest measured density is to reduce the dielectric screening to $\ve \lesssim 10^3$. This reduction may result from local hardening of the soft mode induced by the doping sites~\cite{bauerle1980soft,Crandles1999,vanderMarel2008}.

We also find that upon rasing the density the Lifshitz transitions observed by Ref.~\cite{Lin20133} have a weak effect on $T_c$. The interaction between the plasmon and the longitudinal optical phonon has a much stronger effect and leads to a suppression of $T_c$ at high density (see Fig.~\ref{fig:domes}). We also show that the plasmon mechanism for superconductivity can explain the observed transition temperatures in two-dimensional gases based on SrTiO$_3$~\cite{caviglia2008electric,ueno2008electric}.
Finally, we show that the plasmon leads to a density dependant feature in the tunneling density of states (see Fig.~\ref{fig:TDOS}).

\noindent
{\it Model --}
For simplicity the three $t_{2g}$ conduction bands near the $\G$-point are taken to be isotropic and parabolic with a dispersion $\e_{\bs k ,a} = {k^2/ 2m_a }-e_a$, where $a=1,2,3$ labels the bands. We take $m_1 \approx2 \;\mrm{to} \; 4 \,m_e$, $m_2 = m_3 = 2 \,m_e$~\cite{Lin20133} and $e_1 = \e_F$, $e_2 = \e_F -\d E_2$ and $e_3 = \e_F -\d E_3$ with $\d E_2 = 2 \, \mrm{meV}$ and $\d E_3 = 8 \, \mrm{meV}$. We start our analysis from the lowest density, where $\e_F < \d E_2$ and only the lowest band is occupied. Therefore all quantities refer to the lowest band unless explicitly specified otherwise.

To describe the interactions between the electrons we consider only long-range Coulomb forces, and use the random-phase-approximation
\be
V(i\w,q) =  {4\pi e^2 \over \ve (i\w) q^2-4\pi e^2\Pi(i\w,q)}\,,\label{V}
\ee
where
$
\Pi(i\w,q)
$
is the electronic polarization and $\ve(i\w)$ is given by Eq.~\ref{e(w)}.

\noindent
{\it plasma osculations in a slightly doped ionic crystal --}
Before estimating the transition temperature from Eq.~\ref{V} we discuss the interaction between the electronic and ionic longitudinal modes.
At long wavelengths the electronic polarization $\Pi(i\w,q)$ leads to a plasma mode $\w_\infty \equiv \sqrt{4 \pi e^2 n / \ve_{\infty} m}$, which hybridizes with the longitudinal mode $\w_L$ (see SI and Refs.~\cite{Mooradian1966,cohen1969superconductivity}).
When $\w_L \gg \w_\infty$ the plasma frequency is reduced to $\w_0 \equiv \sqrt{\ve_\infty / \ve_{0} } \;\w_\infty$ due to screening by the crystal.
%As we will show, in this limit the plasma frequency is smaller than the Fermi energy and may lead to pairing.
On the other hand if $\w_\infty \gg \w_L$ the plasma mode takes its bare value and screens the electric fields induced by the longitudinal mode $\w_L$. As a result the gap between $\w_L$ and $\w_T$ at $q \rightarrow 0$ disappears and the phonon mode decouples from the electrons~\cite{mahan2013many}.
%Here the plasma frequency is larger than $\w_L$ and is therefore definitely too large to lead to pairing.

\begin{figure}
\centering
\includegraphics[width=0.9\linewidth]{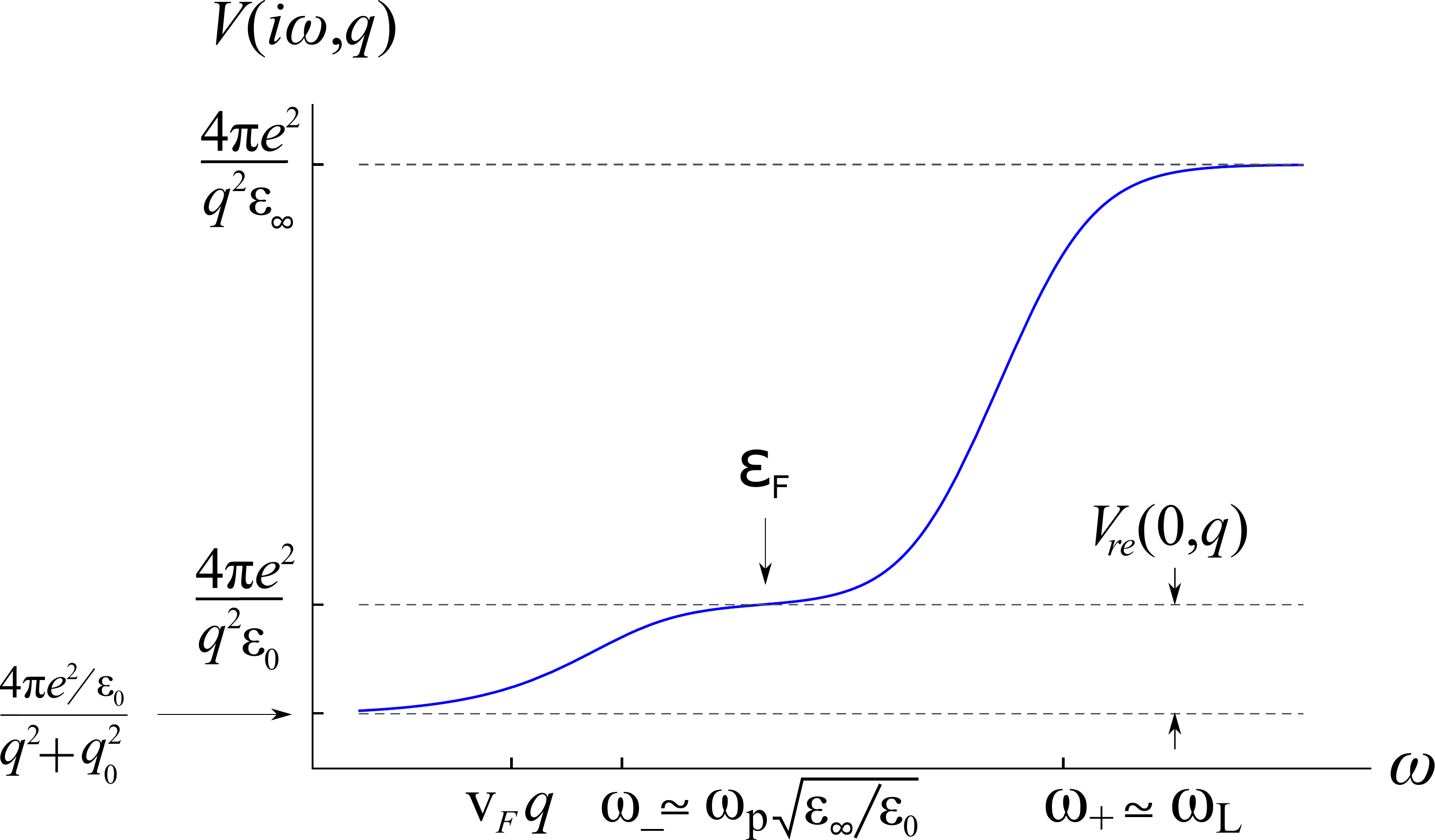}
\caption{ Schematic plot of the interaction Eq.~\ref{V} for $\w_\infty \ll \w_L$ showing two attractive contributions (see text). Here $q_0 = \sqrt{4\pi e^2 \nu /\ve_0 }$ is the screened Thomas-Fermi wavelength.}\label{fig:V}
\end{figure}

Both of these limits are realized in doped strontium titanate.
In what follows we focus low density, {\it i.e.} $n_{3d} \sim 10^{17} \,\,\mrm{to}\,\, 10^{19}$ cm$^{-3}$ where the plasma frequency is lower than $\w_L$ and therefore there is a small plasma mode lying below the Fermi energy.
Fig.~\ref{fig:V} presents the interaction Eq.~\ref{V} in this limit.
As can be seen the interaction is essentially frequency independent in the vicinity of $\e_F$ and it is physically obvious that the frequency dependence at the scale of $\w_L$ cannot give rise to pairing.
Nevertheless, all earlier studies~\cite{Koonce1967,Appel1969,Eagels1969,Takada1980,Klimin2014} assume that Eliashberg theory continues to hold and integrate the interaction up to many times $\e_F$ ($10^4 \,\e_F$ in the case of Ref.~\cite{Takada1980}) to obtain $T_c$. The problem is that electronic states far above the Fermi level are involved, and there is no reason to single out the Eliashberg paring diagrams as the dominant ones. This problem has been emphasized by Ref.~\cite{Grabowski1984} which showed that inclusion of vertex corrections rapidly kill $T_c$ once the frequency of the bosons that are being exchanged become  comparable to $\e_F$. This work explains why previous proposals~\cite{Takada1978,Rietschel1983} of the plasmon exchange mechanism in metals are not valid, because otherwise very high transition temperatures are predicted. Form this point of view the novelty of SrTiO$_3$ is that due to crystal screening the plasmon is weakly coupled and can be smaller than $\e_F$.

Our departure from previous work is to insist that when the energy scale is much lower than $\w_L$, we live in a world where the bare Coulomb repulsion $e^2/\ve_{\infty}r$ is replaced by $e^2/\ve_{0}r$, which sets the strength of the interaction. We therefore restrict our frequency integration to $\e_F$ and below when we solve the Eliashberg equation. For $\w \gg v_F q$ this leads to the interaction
\begin{align}
V(i\w,q) \approx {q_{0}^2 \over\nu q^2}\left[ 1- {\w_0 ^2 \over \w_0^2 +\w^2} \right]\,.\label{V2}
\end{align}
where $q_{0} \equiv \sqrt{4 \pi e^2 \nu / \ve_{0}}$ is the screened Thomas-Fermi wavelength and $\nu \equiv m k_F/ \pi^2$ is the density of states of the lowest conduction band.
To relate to standard Eliashberg theory we decompose the interaction into two parts: a static repulsive part $V_{st}(q)\equiv  { q_{0}^2 \over \nu q^2}$ and the retarded attractive piece $V_{re}(i\w,q) \equiv V_{st}(q)-V(i\w,q)$.

Considering the interaction in Eq.~\ref{V2} as a source for superconductivity we immediately encounter a problem. If we assume that slightly doped SrTiO$_3$ has the same $\ve_0\approx 2\times 10^4$ as undoped SrTiO$_3$, the effective coupling strength $\lambda \sim q_0^2 / 2k_F^2$ is of order $10^{-2}$ and there is no hope of getting any measurable $T_c$. This forces us to use $\ve_0$ as a phenomenological parameter (and therefore $\w_T$, if we continue to assume the validity of the LST relation) and see what we need to set a measurable $T_c$. We do this by considering dependance of the transverse frequency $\w_T$ on doping
\be
\w_T^{-1} (n) = \w_{sat}^{-1}+{ \w_T^{-1}(0) - \w_{sat}^{-1} \over \sqrt{1+{\a n}}}\,, \label{wT}
 \ee
 where $\w_T(0) = 1.9\, \mrm{meV}$~\cite{Vogt1995,vanderMarel2008}. $\a$ and $\w_{sat}$ control the onset density and the high density saturation frequency, respectively. The reduction of $\ve_0$ is obtained from Eq.~\ref{wT} through the LST relation $\ve_0 =\ve_\infty \left( \w_L / \w_T \right)^2$. As we see below, at the lowest density we need $\ve_0 \approx 10^3$. { At the end of the paper we speculate how local stiffening of $\w_T$~\cite{bauerle1980soft,Crandles1999,vanderMarel2008} can lead to suppression of $\ve_0$ in the vicinity of the doping sites.}

\noindent
{\it Eliashberg theory -- }
To solve for the superconducting gap we employ the Eliashberg theory~\cite{Eliashberg1960}. For brevity we do not derive the self-consistent equations (for a review see Ref.~\cite{Margine2013}). We consider all three self-consistent equations for the the case in which the gap has $s$-wave symmetry, which are given by
\begin{align}
&\phi(i\w, k) =- {T\over N}\sum_{\w', k'}{\phi (i\w', k') \over A(i\w', k')}\langle V \left( i\w-i\w',q\right) \rangle
\label{phi}
\\
&\chi(i\w , k) = -{ T\over N } \sum_{\w', k'} {\e_{ k'}+\chi(i\w', k')\over A(i\w', k')} \langle V_{re} \left( i\w-i\w',q\right) \rangle \nn
\\
&Z(i\w, k ) = 1+{ T\over \w \,N } \sum_{\w', k'} {\w'Z(i\w', k')\over A(i\w', k')} \langle V_{re} \left( i\w-i\w',q\right) \rangle \nn
\end{align}
where $q \equiv |\bs k - \bs k'|$ and  $A(i\w, k)\equiv \left[Z(i\w, k)\,\w\right]^2+\phi^2(i\w, k)+\left[ \e_{k} + \chi(i\w, k) \right]^2$.
Here $Z(i\w, k)$, $\chi(i\w, k)$ and $\phi(i\w, k)$ represent the mass renormalization, the dispersion renormalization and the superconducting order-parameter appearing in the self-energy corrections $\Sigma(i\w , k) = \left[1-Z(i\w ,k)\right]\,i\w \,\s^0 + \chi(i\w , k)\, \s^3 +\phi(i\w , k)\, \s^1$ to the Green's function
\be
G(i\w ,\bs k) = \left[ G_0 ^{-1} (i\w , k)-\Sigma(i\w , k)\right]^{-1}\,,\label{G}
\ee
where $G_0 = \left( i\w-\e_k\,\s^3\right)^{-1}$ and the Pauli matrices $\s^i$ act in Nambu space $\psi^\dag _{\bs k}=(c_{\bs k\da}^\dag,c_{-\bs k}\ua)$.

The brackets in Eq.~(\ref{phi}) denote averaging over the solid angle $\langle V(i\w,q) \rangle \equiv \int_{-1}^1 {dl\over 2\pi}\; V\left(i\w,\sqrt{k^2 + k'^2 -2 k k'\, l} \right)$ such that $\langle V_{st}(q)\rangle = { q_{0}^2\over 2\pi\nu k k' }\log \left|  {k+k' \over k-k'}\right| $.
The angular integration over the retarded part of the interaction is cutoff at large angles when $q > \w/v_F $ where the interaction in Eq.~\ref{V} becomes statically screened (see Fig.\ref{fig:V}). As a result the height of the Lorentzian in Eq.~\ref{V2} is reduced compared to the static part $V_{st}(q)$ by
\[\langle V_{st}(q)  \rangle-\langle V_{re}(0,q) \rangle  =  {  q_{0}^2\over 4\pi \nu k k'} \log \left[{(k+k')^2 + q_{0}^2 \over  (k-k')^2 + q_{0}^2}\right]\,,\]
which is nothing but the solid angle average of Eq.~\ref{V} in the limit of $\w\rightarrow 0$.

To solve the Eliashberg equations Eq.~\ref{phi} numerically we truncate the sum over Matsubara frequencies by setting a cutoff frequency $\Omega = 4 \w_0$. In conventional Eliashberg theory the renormalization of the static Coulomb interaction $V_{st}(q)$ due to integration over frequencies higher than the cutoff is taken into account by the phenomenological Coulomb pseudo-potential $\mu^*$. Here we will need to introduce a similar phenomenological parameter, which is a dimensionless ratio $\eta<1$
\be
V \left( i\w,q\right)  = \eta V_{st}(q) - V_{re}(i\w,q)\,.
\ee
Note that the conventional $\mu^*$ is related to $\eta$ through the double momentum average $\mu^* = \eta \nu \langle \langle V_{re}(q)\rangle \rangle $ on the Fermi surface~\cite{Margine2013}.
The solution of Eq.~\ref{phi} is then obtained by iteration of the equations starting from the initial state $Z(i\w,k) = 1$, $\chi(i\w,k) = 0$ and $\phi(i\w,k) = \phi_0$ if $|\w| < \w_0 $ and zero otherwise.

The momentum dependence of the solutions of Eq.~\ref{phi} strongly depends on the coupling strength $ \lambda \equiv  q_{0}^2 / 2 k_F^2$ (see SI). For strong coupling the order parameter $\phi(i\w,k)$ extends far away from the Fermi surface. However, at weak coupling it becomes sharply peaked, signaling that most of the pairing occurs in a narrow window around $k = k_F$. Therefore, we further simplify Eq.~\ref{phi} by restricting the momentum integration to the vicinity of the  Fermi surface by integrating the strong momentum dependence coming from the dispersion in $A(i\w,k)$ and from the Coulomb interaction while setting to $k' = k_F$ in all other quantities. In this limit the dispersion renormalization $\chi(i\w,k)$ also becomes much smaller than $\e_F$ and can be neglected (see SI). We emphasize that this procedure is valid only at weak coupling.

The calculated transition temperature is plotted in Fig.~\ref{fig:domes}.a for two different sets of parameters. The blue curve corresponds to $\eta = 0.3$, $m_1 = 4 m_e$, $\a = 8\times 10^{-18}cm^{3}$ and $\w_{sat} = 18\, \mrm{meV}$, and cyan to $\eta = 0.4$, $m_1 = 2 m_e$, $\a = 5.5\times 10^{-16}cm^{3}$ and $\w_{sat} = 11.5\, \mrm{meV}$.
Here the higher density dome is taken with a higher mass due to the mass enhancement measured by Ref.~\cite{Lin20133}.
We also plot the plasma frequency $\w_0$, the Fermi energy $\e_F$ and the frequency of the transverse mode $\w_T$ for each one of these sets in Fig.~\ref{fig:domes}.b and Fig.~\ref{fig:domes}.c.
The transition temperature is compared with the experimental data points (grey) taken from Refs.\cite{Schooley1964,Lin20133}.

The reduction of $T_c$ at higher doping occurs because the plasma frequency $\w_\infty$ grows and becomes comparable to $\w_L$, where the two modes hybridize. In this limit the electron gas begins to screen to the crystal fields and not vice versa, which leads to a decoupling of the longitudinal optical mode from the electrons (see SI).
Therefore, the plasmon mechanism cannot explain superconductivity in the high density regime $n_{3d} \sim 10^{19}\;\mrm{to}\;10^{21}\mrm{cm}^{-3}$.
%However, in this regime the Fermi energy is higher and therefore there are alternative mechanisms that may explain superconductivity. For example the lowest longitudinal mode is found to have a moderate coupling strength to the electrons~\cite{Meevasana2010}.

\noindent
{\it Superconductivity in two-dimensions  --}
\begin{figure}
\centering
\includegraphics[width=8.5 cm,height=5cm]{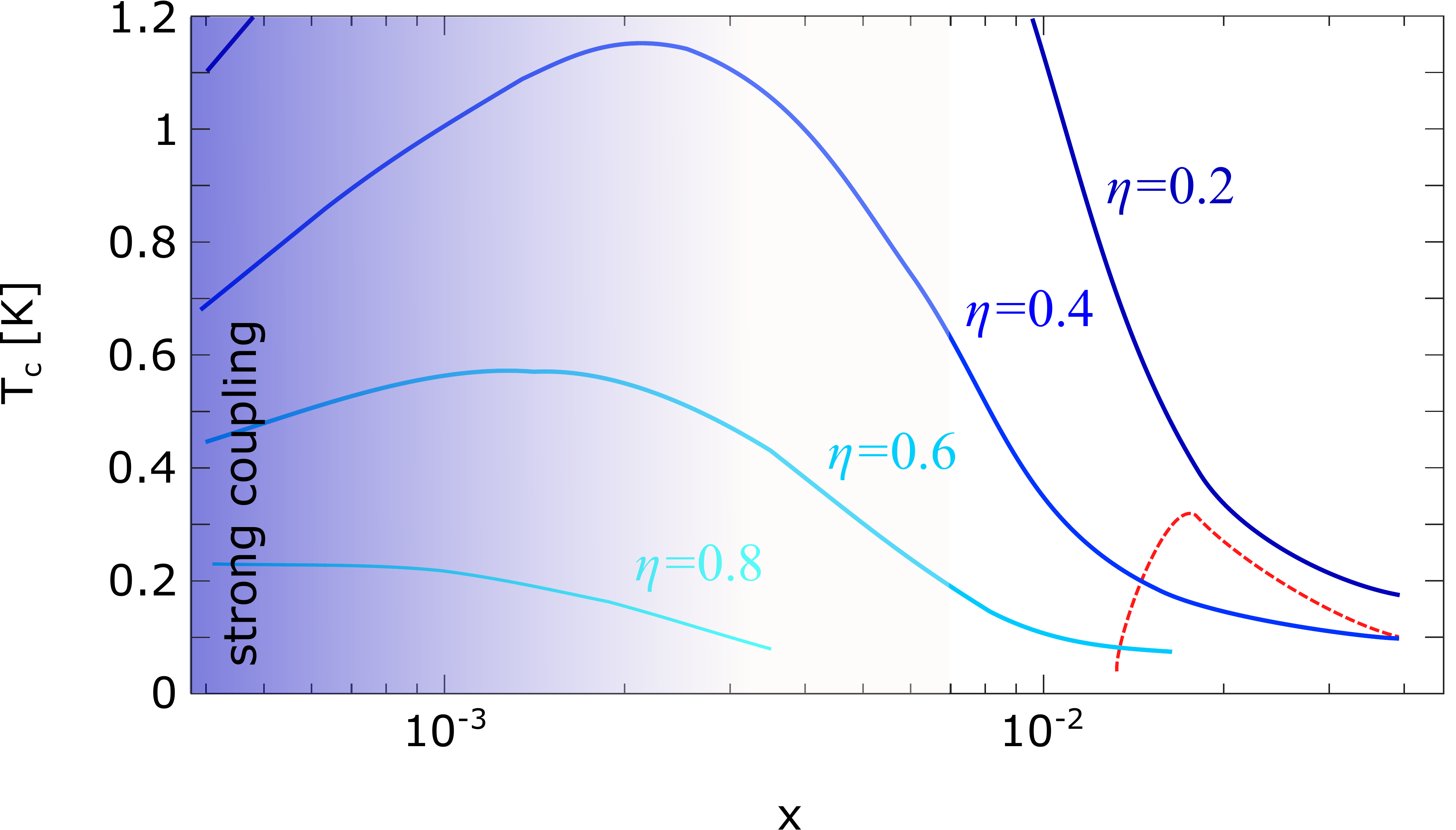}
\caption{The transition temperature in two-dimensions vs. doping for $\w_T = 15\,\mrm{meV}$ and $m = m_e$ and for $\eta = 0.2,\,0.4,\,0.6$ and $0.8$. The red dashed line is the typical $T_c$ from experiments~\cite{richter2013interface}. The region of strong coupling where the calculation is not valid is shaded.   }\label{fig:2d}
\end{figure}
A variety of two-dimensional electron gases have been realized in SrTiO$_3$ (for example Refs. \cite{ohtomo2004high,santander2011two}). These gases become superconducting with a transition temperature which is similar to the bulk~\cite{ueno2008electric,caviglia2008electric}. Therefore, it is interesting to understand whether the plasmon mechanism described here is relevant in two-dimensions. We note that in this case the observed superconductivity is limited to rather high doping levels (see Fig.~\ref{fig:domes}.a), where $\e_F \sim 50 \,- \,100 \,\mrm{meV}$,

To address this question we repeat the derivation of the Eliashberg equations for case of two-dimensions (see SI).
A crucial difference is that now the plasma frequency is not gapped $\w_\infty(q) =  \sqrt {2\pi e^2 n q / \ve_\infty m}$.
This leads to a small attractive interaction below the Fermi energy even when the doping level is high.
Additionally, the reduction of $\ve_0$ (or the stiffining of $\w_T$) are much more natural due to the electric fields near the surface. Since there are no systematic measurements, we simply assume a constant $\w_T = 16\,\mrm{meV} $, which corresponds to $\ve_0 \approx 180$.

The calculated transition temperature vs. density for a single band model with $m = m_e$ is plotted in Fig.~\ref{fig:2d} for different values of $\eta$.
Note that here we have taken also smaller values of $\eta$ because the Fermi energy is larger and the traditional $\mu^*$ renormalization applies (see SI). The shaded region is where the coupling strength becomes large and the Eliashberg theory should not be trusted. The red dashed line is a typical $T_c$ dome from experiments~\cite{caviglia2008electric,richter2013interface}.

As can be seen at lower density the calculated $T_c$ dome does not agree with experiment, and extends to very low density where the localization is observed~\cite{caviglia2008electric}. This discrepancy is actually consistent with the findings of Ref.~\cite{richter2013interface}, which report pseudogap behavior in this regime. Therefore, the reduction of $T_c$ at lower density results from phase fluctuations and not decreasing of the pairing gap, such that the mean-field $T_c$ is much higher than the observed one.
%Indeed, disorder becomes important at relatively high doping~\cite{Bell2009}, which is expected to suppresses the superfluid density and enhance phase fluctuations.

\noindent
{\it Tunneling density of states -- }
We now turn to discuss a feature of plasmonic superconductivity which show in the single particle tunneling density of states (TDOS) above the gap.
The TDOS of a standard BCS superconductor displays fingerprints of phonon resonances~\cite{Schrieffer1963}. As we show here, and for the same reason, the plasma frequency in dilute SrTiO$_3$ should also become observable.

We obtain the TDOS from the imaginary part of the analytically continued Green's function (Eq.~\ref{G}), which is calculated using a controlled Pad\'e approximation (for details see Ref.~\cite{Beach2000} and SI).
In Fig.~\ref{fig:TDOS} we plot the TDOS for different values of $n_{3d}$ ranging between $5\times 10^{17}\,\mrm{cm}^{-3}$ and $5\times 10^{18}\,\mrm{cm}^{-3}$ at a temperature $T = 30 \, \mrm{mK}$ and using $\eta = 0.5$, $m = 2m_e$, $\a = 5\times 10^{-16}\,\mrm{cm}^3$ and $\w_{sat} = 10 \,\mrm{meV}$. The spectral line-shape of the plasmon exhibits strong density dependence which is not expected in the case of phonon mediated superconductivity.

\noindent
{\it Discussion -- }
We claim that the plasmon is the only bosonic mode capable of explaining superconductivity in dilute SrTIO$_3$. On the other hand we need to assume reduction of crystal screening by considering the stiffening of the soft ferroelectric mode induced by defects.
The defects may result from oxygen vacancies~\cite{bauerle1980soft,Crandles1999} or from chemical dopants (Nb, La, etc.)~\cite{vanderMarel2008}.
These defects induce pinning potentials and long range distortions, and it is known that $\omega_T$ is highly sensitive to strain induced by pressure or stress due to the proximity of the ferroelectric transition~\cite{Venturini2003,worlock1967electric}.
The oxygen vacancies are expected to have a stronger effect than substitutional disorder.
We account for this difference by using different onset densities for the stiffening which leads to the two domes in  Fig.~\ref{fig:domes}.
In this scenario the two domes observed by Ref.~\cite{Lin20133} are related to different doping techniques and possibly differences in the sample properties (for example in Ref.~\cite{de2015interplay} the relatively low value of $\ve_0 \approx 5\times 10^3$ measured in their pristine SrTiO$_3$) rather than the Lifshitz transitions. It is also important to emphasize that the values of $\w_T$ presented in Fig.~\ref{fig:domes} have been inferred from the LST relation, which may breakdown due to disorder.

The plasmon mechanism cannot explain superconductivity at higher density regime, $n\sim 10^{19}\;\mrm{to}\;10^{20} \, \mrm{cm}^{-3}$, (see  Fig.~\ref{fig:domes}) because the plasma frequency becomes larger than $\w_L$. Interestingly, Ref.~\cite{Meevasana2010} estimated the electron-phonon coupling strength to the lowest longitudinal mode $\w_{L1} = 21\,\mrm{meV}$, and found it to be moderate. It is therefore possible that this mode is mainly responsible for the pairing at higher densities.

An interesting aspect of the plasmonic mechanism in two-dimensions is that it is expected to be highly sensitive to external metallic gates. If a high density metal is deposited close enough to the two-dimensional superconductor the long-ranged Coulomb potentials will be screened, which should dramatically modify the dispersion of the plasma mode. Therefore, it is interesting to understand the effects of external metallic gates on the superconducting states in two-dimensions.

Finally, It is compelling to understand whether the plasmonic mechanism is relevant to other materials? According to our predictions the important ingredients are a dilute electron gas with relatively large effective mass and strong dielectric screening such that the plasma frequency lies below the Fermi level. The recently discovered superconductors in doped topological insulators~\cite{Hor2010,wray2010observation,kriener2011bulk,liu2015turning} are candidates which may match these criterions, where the large dielectric screening is naturally present due to the small band gap.

{\it Acknowledgments -- }
We are grateful to Leonid Levitov for insightful discussions which helped launch this project and to Nandini Trivedi and Mohit Randeria for helpful discussions. JR acknowledges the Gordon and Betty Moore Foundation under the EPiQS initiative under grant no. GBMF4303. PAL acknowledges the support of DOE  under grant no. FG02-03ER46076.

\begin{figure}
\centering
\includegraphics[width=0.9\linewidth]{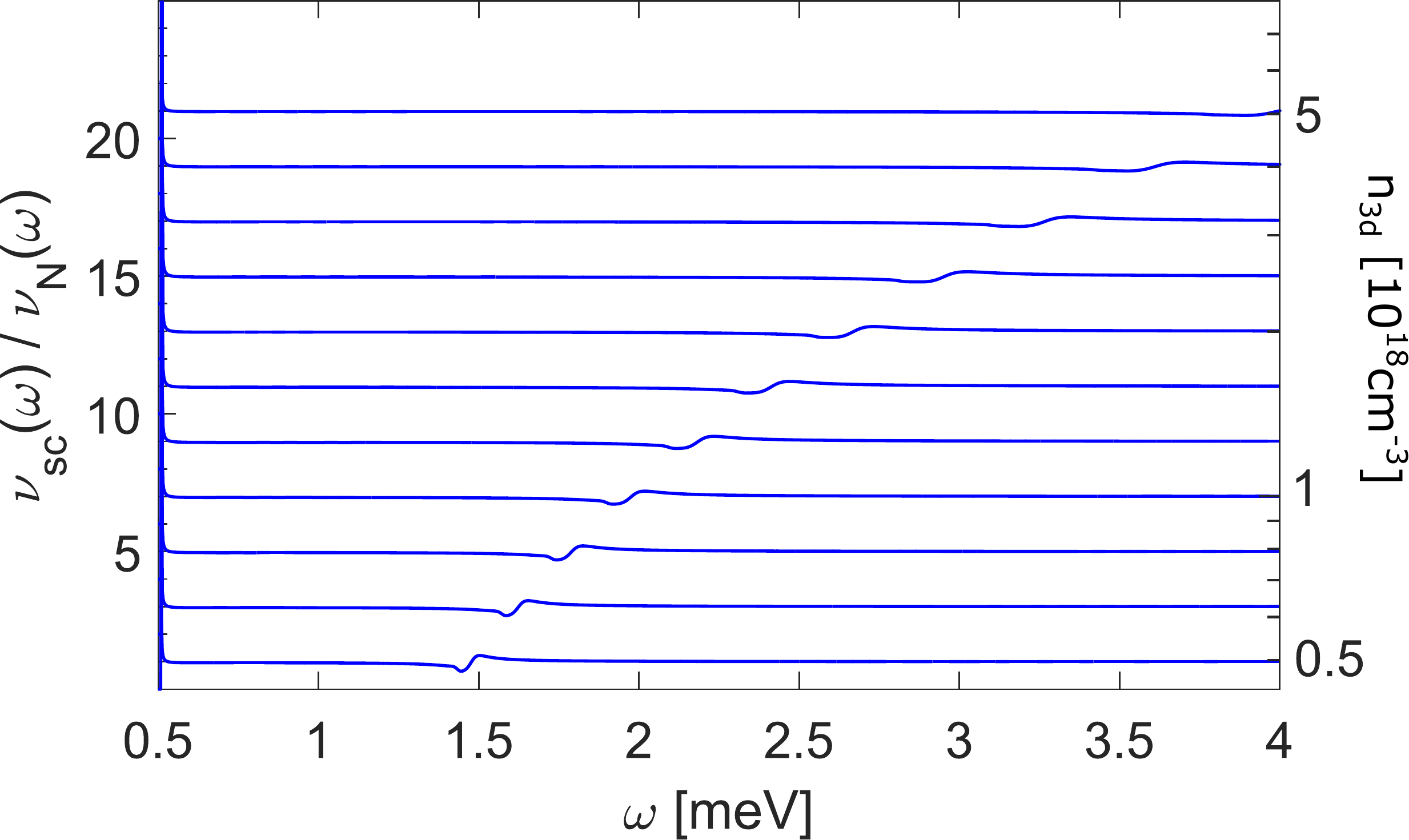}
\caption{The superconducting TDOS $\nu_{sc}(\omega)$ divided by the normal TDOS $\nu_N(\w) $ for different values of the density calculated from the analytic continuation of Eq.~\ref{G}. The curves are shifted from each other by $2$ to make them distinguishable. The resonant feature appearing around $\w = \w_0$ depends strongly on the electronic density and is therefore a 'finger print' of plasmonic superconductivity. The gap is barely observable on this scale. }\label{fig:TDOS}
\end{figure}

\onecolumngrid

{\center{\Large{\bf  Superconductivity at very low density: the case of strontium titanate - Supplementary Information}}}

\section{Optical phonon spectrum in $\mrm{SrTiO}_3$}
In this section we review the dielectric properties of insulating strontium titanate including three active optical modes and discuss the applicability of the single resonance model.  We also explain our disagreement with the screened interaction used in a recent preprint (Ref.~\cite{gor2015phonon}).

The dielectric function comes from the sum of the polarizability of various transverse modes $\w_{Tj}$
\be
\ve(i\w)  = 1 +  {4\pi\bs P \over \bs E} = \ve_\infty + \sum_{j}z_j\, {\w_{Tj}^2 \over \w_{Tj}^2 + \w^2} \label{e(w)_SI_0}
\ee
In the experimental literature the crystal dielectric constant is traditionally presented as a product
\be
\ve(i\w) = \ve_\infty  \prod_{j=1} {\w_{Lj}^2 + \w^2 \over \w_{Tj}^2 +\w^2}\label{e(w)_SI}
\ee
which makes it explicit that $\w_{Lj}$ and $\w_{Tj}$ are the zeros and poles of the dielectric constant, respectively. The $\w \rightarrow 0$ limit leads to the Lyddane-Sachs-Teller relation.

The experimentally measured values of the longitudinal and transverse frequencies in insulating SrTiO$_3$ are given by~\cite{Vogt1981,Kamaras1995} $\w_{T1}\approx1.9\,\mrm{meV},\;\w_{T2}=21.2\,\mrm{meV},\;\w_{T3}=68\,\mrm{meV},\;\w_{L1}=21\,\mrm{meV},\;\w_{L2}=59\,\mrm{meV},\;\w_{L3}=98\,\mrm{meV}$ and $\ve_\infty = 5.1$.

\begin{figure}
\centering
\includegraphics[width=8cm,height=4.5cm]{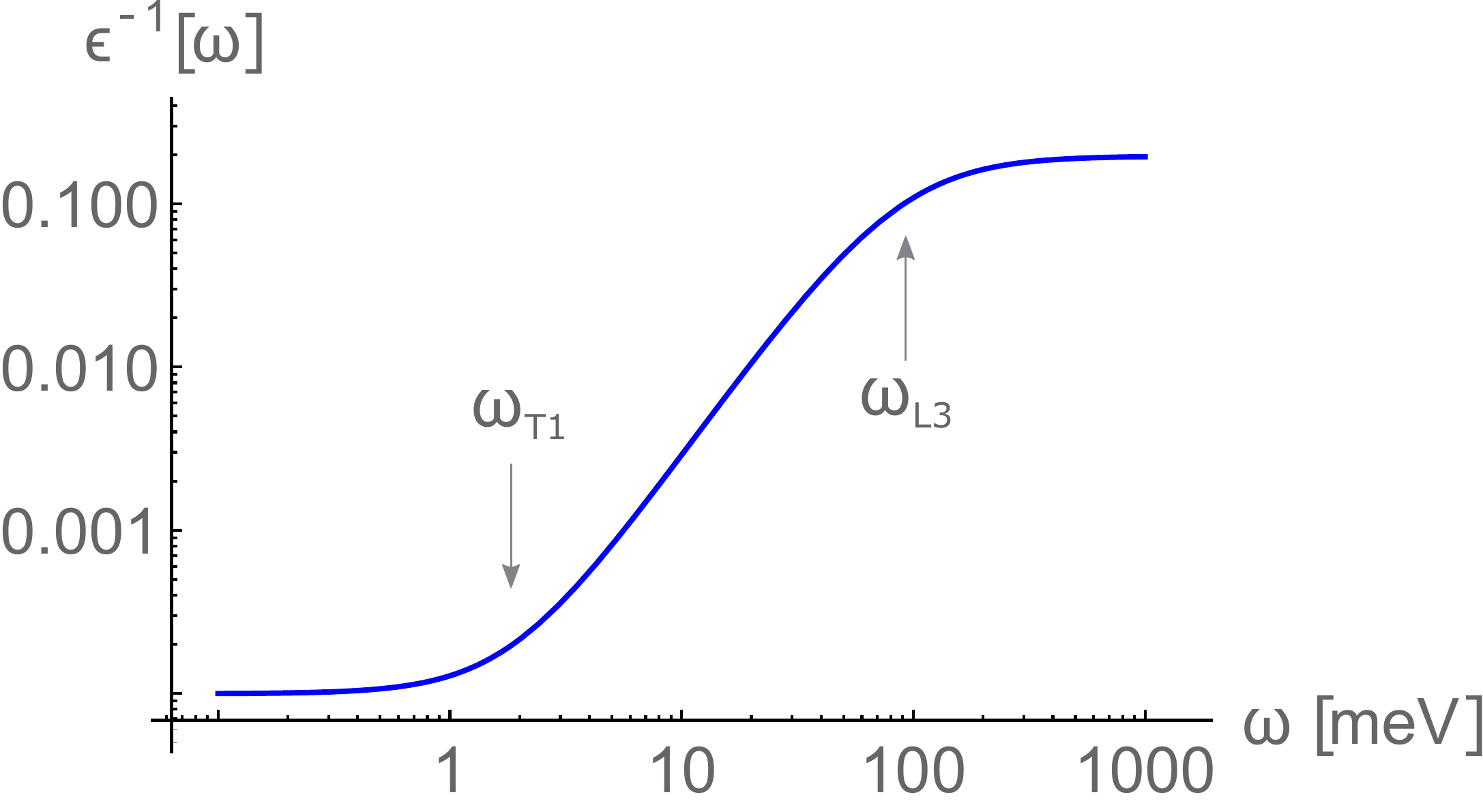}
\caption{The dielectric function Eq.~\ref{e(w)_SI} in insulating SrTiO$_3$ as a function of Matsubara frequency.   }\label{fig:dielectric}
\end{figure}

It is convenient to decompose the inverse dielectric constant (or equivalently the screened Coulomb interaction) into a sum of resonances
\be
{1 \over \ve(i\w)  } = {1\over \ve_{\infty}  } \left [ 1 - \sum_{j = 1} \g_j {\w_{Lj}^2 \over \w_{Lj}^2 +\w^2}\right] \label{e(w)_SI_2}
\ee
Eq.~\ref{e(w)_SI_2} has the interpretation that the screened interaction $V_{sc} = 4\pi e^2 / q^2\ve(i\w)$ is given by the bare Coulomb repulsion reduced by the contributions from each $\w_{Lj}$ modes.
Using the experimentally measured values for the parameters in Eq.~\ref{e(w)_SI} one gets $\g_1 < 0.001 $, $\g_2 \approx 0.183$ and $\g_3 \approx 0.815$~\cite{eagles1965polaron}. Thus we find that that the coupling to the lowest mode $\w_{L1}$ and $\w_{L2}$ is weak.
This is also seen from Eq.~\ref{e(w)_SI} by noting that $\w_{T2} \approx \w_{L1}$ and $\w_{T3} \approx \w_{L2}$ so that their contribution to Eq.~\ref{e(w)_SI} cancel each other, leaving only $\ve_{\infty} \left(\w_{L3}^2 + \w^2 \right)/\left(\w_{T1}^2 + \w^2 \right) $.
This is consistent with the conclusions of Ref.~\cite{Zhong1994} which find that the lowest transverse mode is most closely related to the largest longitudinal mode $\w_{L3}$. Therefore we can use the single resonance model
\be
\e(i\w) \approx \ve_{\infty} +{(\ve_{0}-\ve_{\infty})\w_{T1}^2 \over \w_{T1}^2+\w^2 }
\ee
where $\e_0 \approx 2 \times 10^{4} \approx \w_{L3}^2 / \w_{T1}^2$. We therefore neglect the index j in the main text and identify $\w_T = \w_{T1}$ and $\w_{L} = \w_{L3}$. The full dielectric constant, including the contributions from 3 modes, is plotted in Fig.~\ref{fig:dielectric} which indeed resembles a wide single resonance with transverse frequency $\w_{T1}$ and longitudinal frequency $\w_{L3}$.

It is however interesting to point out that Ref.~\cite{Meevasana2010} estimated a moderate coupling to the first longitudinal mode, which is much stronger than the values given earlier after Eq.~\ref{e(w)_SI_2}~\cite{eagles1965polaron}.
This estimate is based on the self-energy correction to the electron dispersion measured by photoemission and involves involves a finite value of $q$, whereas the optical measurement of $\g_i$ are for $q= 0$ thus it is possible that coupling is stronger at finite $q$ and it could be that the mode at $\w_{L1}$ plays an important role and that at higher density, where the chemical potential is higher than $\w_{L1}$ it also participates in the pairing. In any case we are mostly interested in the case where the Fermi energy is significantly lower than $\w_{L1}$.

We remark that from Eq.~\ref{e(w)_SI_0} and Eq.~\ref{e(w)_SI} it is clear that $\ve(i\w)$ is always positive, i.e. the Coulomb interaction screened by polar phonons is always positive when the frequencies are expressed in Matsubara space. In particular, the static interaction is always repulsive.
This feature however is not explicit in Eq.~\ref{e(w)_SI_2}. Nevertheless, there must be constraints on the parameters $\g_j$ so that Eq.~\ref{e(w)_SI_2} also satisfies the positivity requirement, i.e. $\g_j$ cannot be chosen as free parameters.

A recent paper by Gor'kov~\cite{gor2015phonon} appears to have fallen prey to this pitfall. He argues that in SrTiO$_3$ the mode $\w_{L3}$ is mainly responsible for the large $\ve_0$ and he introduces (as we do) the standard form of the interaction in the case of a single phonon resonance (his Eq.~1)
\be
V(i\w,q) = {4\pi e^2 \over q^2 \ve_\infty} -{{4\pi e^2 \over q^2 } \left( {1\over \ve_\infty} - {1\over \ve_0}\right)}{\w_{L}^2 \over \w_{L}^2+ \w^2}\,.\label{gor1}
\ee
As mentioned the last term on the R.H.S. of Eq.~\ref{gor1} may be interpreted as the phonon mediated interaction and corresponds to our Eq.~\ref{e(w)_SI_2} with a single mode.
Gor'kov then argues that since the static repulsion ${4\pi e^2 \over q^2 \ve_0}$ is very small it can be overcome by contributions from $\w_{L1}$ and $\w_{L2}$ if he added their contributions to Eq.~\ref{gor1} as in Eq.~\ref{e(w)_SI_2}. The problem with this argument is that from Eq.~\ref{e(w)_SI_2} and Eq.~\ref{e(w)_SI}, $\w_{L1}$ and $\w_{L2}$ also contributed to $\ve_0$ and adding their contributions again will be double counting. Furthermore, as we discussed earlier, their coupling strengths $\g_1$ and $\g_2$ are not arbitrary, but subject to the constraint $\ve(i\w)$ is always positive. For this reason we disagree with his conclusion that a static attractive interaction can be achieved by polar phonon screening.

\section{interaction between optical phonons and free charge carriers}
In the main text we have briefly discussed the interplay between the plasma oscillations and the low-$q$ behavior of the optical modes in a polar crystal. We have also discussed the behavior at two limiting cases, namely in the limit $\w_\infty \ll \w_L$, where the longitudinal oscillations are much faster than the plasma and therefore simply screen the interaction and modify $\w_\infty$ to $\w_0$. On the other hand in the opposite limit, where  $\w_\infty \gg \w_L$, the plasma is fast enough to completely screen the long range forces induced by the longitudinal mode and therefore the coulomb deriven gap between the longitudinal and transverse modes disappears.
In this section we will make this discussion formal and describe the behavior over the whole range including the hybridization region.

 For this purpose we consider the interaction (Eq.~2 in the main text) in the limit where $\w \gg v_F q$. In this case the polarization bubble can be approximated by $\Pi(i\w ,q) \approx - {3\nu\over 4}\left( {v_F q \over \w}\right)^2+{16\nu\over 5}\left( {v_F q \over \w}\right)^4+\ldots $ and the interaction assumes the form\cite{cohen1969superconductivity,mahan2013many}
\be
V (i\w ,q)\approx {4\pi e^2 \over q^2}{1\over \ve(i\w)/\ve_{\infty} + {\w_\infty^2/\w^2} }  =  {q_{\infty}^2 \over \nu q^2}\left[ 1-(1-\g) {\w_+ ^2 \over \w_+^2  +\w^2}-\g {\w_- ^2 \over \w_-^2 +\w^2} \right] \label{V_3d}
\ee
where $\w_\infty^2= {4 \e_F^2\over 3}\left({q_{\infty}\over k_F}\right)^2 $ is the bare plasma frequency, $q_{\infty} = \sqrt{4\pi e^2 \nu / \ve_{\infty}}$ is the bare Thomas-Fermi momentum,
\be
\w_\pm ^2(q)  =  {\w_L^2 + \w_\infty^2 \over 2} \pm \sqrt{\left({\w_L^2 + \w_\infty^2 \over 2}\right)^2- {\w_\infty ^2 \w_T^2 }}\,,
\ee
\be
\g \equiv  {\w_T ^2 - \w_-^2 \over  \d\w^2}\,,\label{gamma}
\ee
and $\d\w^2\equiv\w^2 _+ - \w^2 _- $.

\begin{figure}
\centering
\includegraphics[width=7cm,height=6.cm]{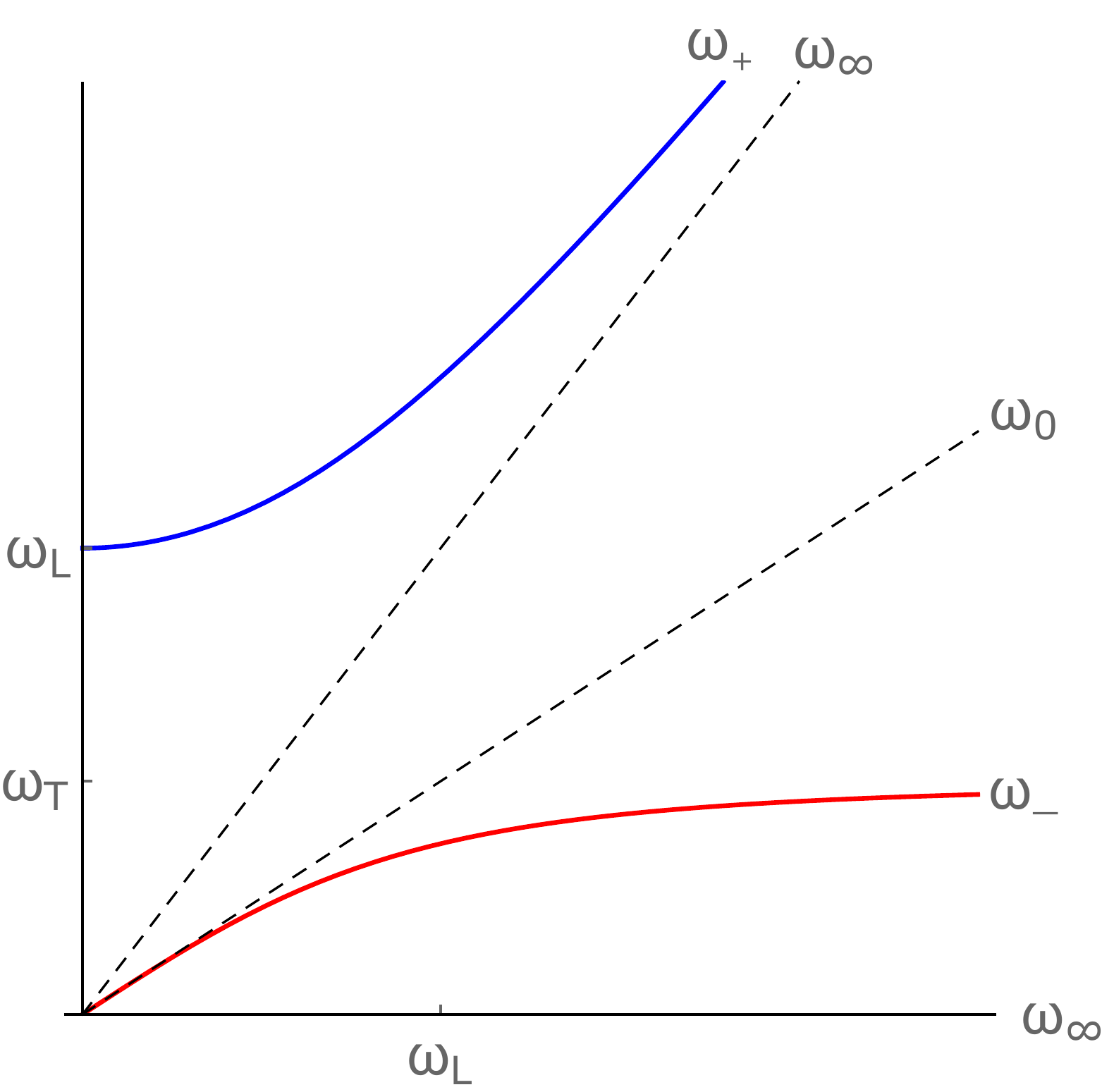}
\caption{The two frequencies $\w_{\pm}$ as a function of the bare plasma frequency. }\label{fig:avoid}
\end{figure}

\begin{figure}
\centering
\includegraphics[width=7cm,height=6.cm]{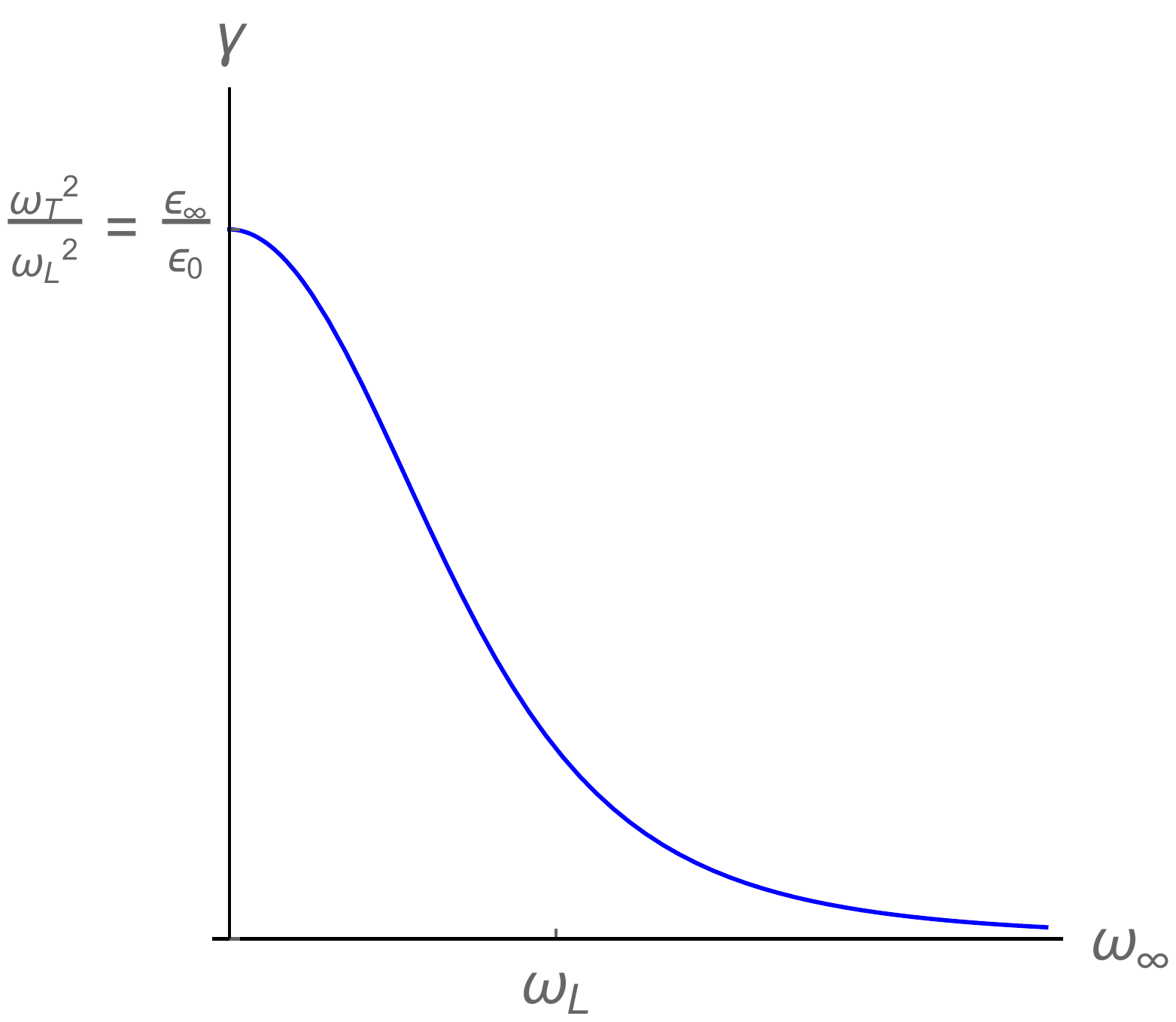}
\caption{The coupling constant $\g$ as a function of the bare plasma frequency. Once the plasma frequency becomes larger than the longitudinal phonon frequency $\w_L$ it completely shields it and the coupling goes to zero.  }\label{fig:gamma}
\end{figure}

The poles of the interaction $\w_{\pm}$ and the coupling constant $\g$ are plotted in Figs. \ref{fig:avoid} and \ref{fig:gamma} as a function of the bare plasma frequency $\w_\infty(0)$.
We identify two distinct regimes: (a) For $\w_\infty \ll \w_L$, we have $\w_+ \approx \w_L$, $\w_- \approx\w_0= \sqrt{\ve_\infty \over \ve_0}\, \w_\infty $ and $\g \approx {\ve_\infty / \ve_0}$. Therefore, in this limit the lower frequency pole corresponds to a plasmon in an interaction which is fully screened by the dielectric.  (b) In the opposite limit, $\w_\infty \gg \w_L$ we have $\w_+ \approx \w_\infty$, $\w_- \approx \w_T$ and $\g \rightarrow 0$. Therefore the plasma frequency takes it's bare value and completely shields the optical phonon.


Doped strontium titanate goes through both of these limits as the density is tuned from $10^{17}\,\mrm{cm}^{-3}$ to $10^{21}\,\mrm{cm}^{-3}$.
Since we are interested in superconductivity at very low density we focus on the case where the bare plasma frequency $\w_\infty\ll \w_L$ and the frequency dependance of the interaction at low energy mainly comes from $\w_-$ (the lower frequency pole). In this case the dielectric constant $\ve _{\infty}$ may be substituted by  $\ve_0 \approx \ve _{\infty}/\g $. It follows that the plasma frequency becomes $\w_\infty \rightarrow \sqrt{\g} \w_\infty = \w_0 $.  We therefore approximate Eq. 2 in the main text by
\be
V(i\w,q) = \g{q_{\infty}^2 \over \nu q^2}\left[ 1- {\w_- ^2 \over \w_-^2 +\w^2} \right] \approx  {q_{0}^2 \over \nu q^2}\left[ 1- {\w_- ^2 \over \w_-^2(q) +\w^2} \right]\label{V_3dapp}
\ee
As the density is increased $\w_\infty$ increases and near $n = 10^{19}\,\mrm{cm}^{-3}$ it becomes comparable to the longitudinal optical frequencies. As can be seen from Fig. \ref{fig:gamma} at that point the coupling $\g$ goes to zero which effects the transition temperature to drop with increasing density.

 \section{Attractive interactions from local couplings to phonons}
In the main text we argued that the plasma osculation is the only relevant resonance below the Fermi energy. One concern that might rise is whether the acoustic phonons are relevant.
In this section we show that at low density the BCS coupling arising from acoustic phonos is negligibly small.

The coupling to longitudinal acoustic phonons arising from local deformations of the lattice is given by\cite{Klimin2014}
\be
{H_{el - ph}^{LA}} =\sum_{k,k'} {(-i q)D \over\sqrt{2 \rho \w_q} }\left(b_q + b^\dag_{-q} \right)c^\dag _k c_{k'}\label{Hel-ph}
\ee
where $\w_q = v_s q$ is the dispersion of the phonons, $v_s = 7.9\times 10^{5} \mrm{cm/s}$ is the speed of sound,  $D= 3 - 5 \, \mrm{eV}$ is the deformation potential and $\rho = 5 \mrm{g/cm^{3}}$ is the mass density. This leads to the following phonon-mediated interaction
\be
V_{LA} (i\w,q) = {D^2 q \over  \rho  v_s} { \w_q\over \w^2 + \w_q^2 }
\ee
Thus the coupling strength can be estimated from the maximum of the Lorentzian times the fermionic density of states per spin
\be
\lambda_{LA} = {\nu D^2 \over 2\rho v_s^2} \label{lam_ac}
\ee
Putting in realistic numbers for $\lambda$ a $n_{3d} = 1\times 10^{18}\, \mrm{cm}^{-3}$ with $m_1 = 2m_e $ and $D = 5\, \mrm{eV}$ one finds that $\lambda \approx 0.01$.

Another possible source for attractive interaction comes from local coupling to the transverse optical mode. Since this mode is transverse it must couple through a vector product. Focusing on low density and therefore projecting this term to the lowest band gives a coupling between the transverse mode and the spin-current in that band~\cite{Kozii2015}
\be
H _{el-ph}^{TO} ={\d t} \sum_{\bs {k q}}c_{\bs{k},s}^\dag\bs u_{\bs q} \cdot\left[ \bs{k }\times \bs \s  \right]_{ss'}c_{\bs {k+q}s'}\label{Hel-ph-TO}
\ee
Note that here we have used the fact that $\D_{so}\gg \e_F$, where $\D_{so}$ is the strength of spin-orbit coupling, and therefore the bands are taken in the eigenstates of spin-orbit coupling. Here $\d t$ is the induced hopping between different orbital due to the transverse distortion $\bs u$.

Following the same procedure as in Eq.~\ref{lam_ac} on obtains an effective coupling strength of
\be
\lambda_{TO} \approx \nu {\d t^2 k_F^2 \over \rho \w_T^2}
\ee
taking an overestimate of $\d t = t = 300 \,\mrm{meV} $ and using the smallest $\w_T$ observed in pristine samples one gets $\lambda_{TO} \approx 5\times 10^{-5} $ at $n = 10^{18}\,\mrm{cm}^{-3}$.

\section{Numerical solution of the Eliashberg equations}
In this section we elaborate on the solution of the Eliashberg equations (Eq.~5 in the main text).
First we discuss the momentum dependent solutions and show that at weak coupling they may be reduced to a simpler {\it isotropic} form and then describe the weak coupling limit restricted to the Fermi surface.

The numerical solution of Eqs.~5 of the main text is obtained by straightforward iteration starting from $\chi = 0$, $Z = 1$ and $\phi = 10^{-4} \e_F$ for $|\w|<\w_-$ and zero otherwise. The integration over momentum is broken into a discrete sum with simple trapezoid rule. We typically used about $60$ grid points per unit $k_F$.
A solution is obtained once the root mean squares defined as follows
\[{\langle|\phi - \phi_0 |\rangle \over \max |\phi_0|}+{\langle |\chi - \chi_0 |\rangle\over \max |\chi_0|}+{\langle |Z - Z_0 |\rangle \over \max |Z_0|}\]
become smaller than $10^-6$, where $\langle...\rangle$ denotes an average over all data points and $\phi_0$, $\chi_0$ and $Z_0$ are the solutions obtained in the previous iteration.
Far from $T = T_c$ a solution is typically obtained after 15 to 20 iterations. As $T_c$ is approached this number diverges (we cutoff after 80 - 100 iterations).

\begin{figure}
\centering
\includegraphics[width=18cm,height=4.3cm]{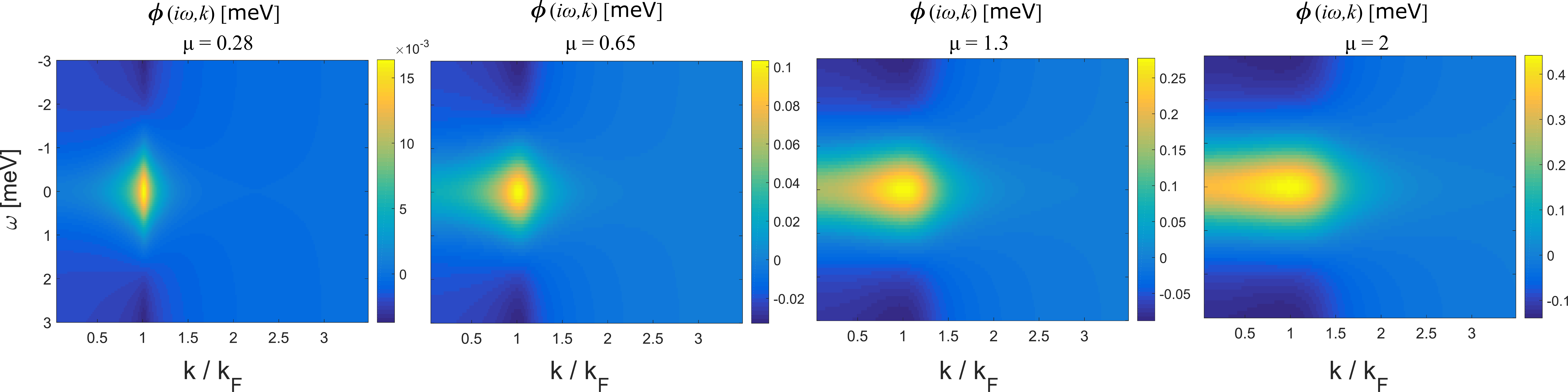}
\caption{An example of the superconducting order parameter $\phi(i\w,k)$ calculated numerically from Eqs.~\ref{phi} for different values of the coupling $\lambda = \g q_{\infty}^2 / k_F^2$ and for $n = 5\times 10^17\,\mrm{cm}^{-3}$, $m = 2m_e$, $\e_F = 1.2\,\mrm{eV}$, $\w_- = 2\e_F/3$, $\eta = 0.2$ and $T = 100\,\mrm{mK}$. As the coupling is reduced the order parameter becomes strongly peaked around the Fermi surface showing that most of the pairing occurs in a narrow window around $k = k_F$.  }\label{phi3d:fig}
\end{figure}

\begin{figure}
\centering
\includegraphics[width=18cm,height=4.3cm]{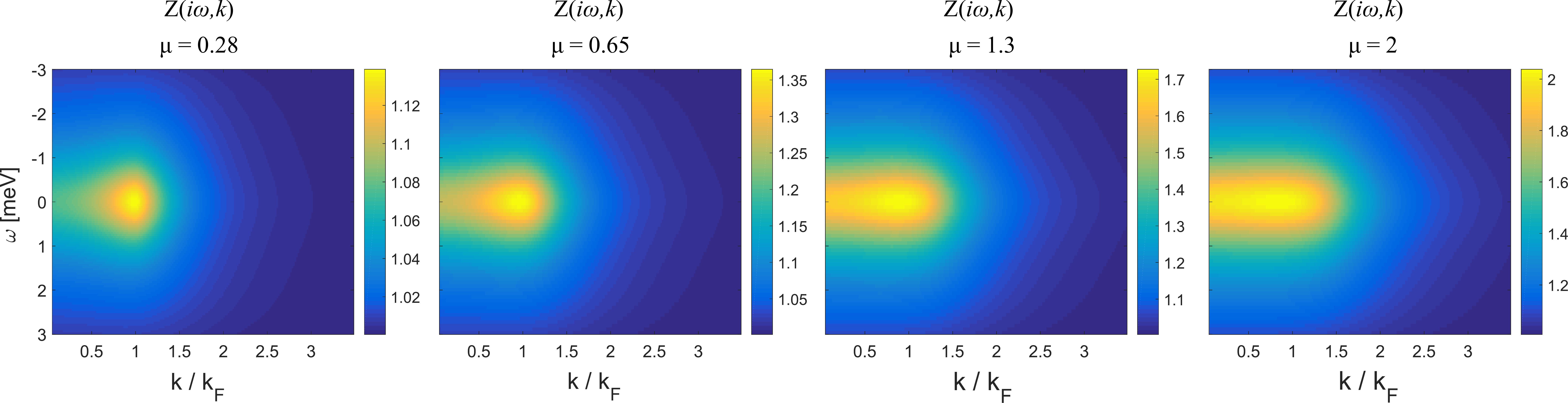}
\caption{An example of the mass renormalization $Z(i\w,k)$ calculated numerically from Eqs.~\ref{phi} for different values of the coupling $\\lambda = \g q_{\infty}^2 / k_F^2$ and for $n = 5\times 10^17\,\mrm{cm}^{-3}$, $m = 2m_e$, $\e_F = 1.2\,\mrm{eV}$, $\w_- = 2\e_F/3$, $\eta = 0.2$ and $T = 100\,\mrm{mK}$.}\label{Z3d:fig}
\end{figure}
\begin{figure}
\centering
\includegraphics[width=18cm,height=4.3cm]{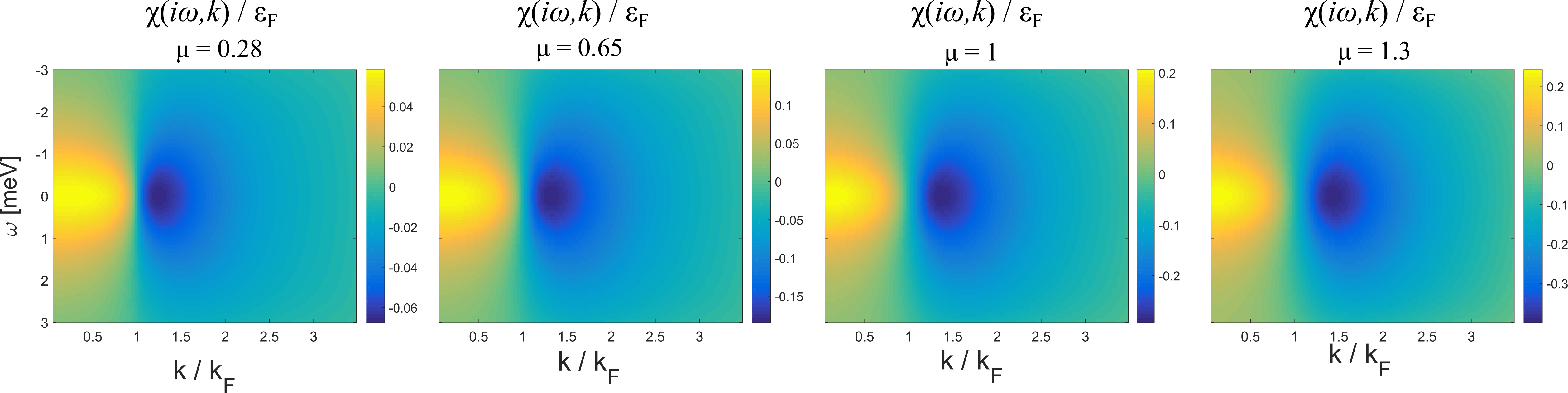}
\caption{An example of the dispersion renormalization $\chi(i\w,k)$ normalized by the Fermi energy calculated numerically from Eqs.~\ref{phi} for different values of the coupling $\lambda = \g q_{\infty}^2 / k_F^2$ and for $n = 5\times 10^17\,\mrm{cm}^{-3}$, $m = 2m_e$, $\e_F = 1.2\,\mrm{eV}$, $\w_- = 2\e_F/3$, $\eta = 0.2$ and $T = 100\,\mrm{mK}$. As the coupling is reduced the ratio between $\chi(i\w,k)$ and the Fermi energy becomes smaller and may be neglected in the weak coupling limit. }\label{chi3d:fig}
\end{figure}

In Figs.~\ref{phi3d:fig} - \ref{chi3d:fig} we plot an example of a solution of Eqs.~\ref{phi} for different values of the coupling strength
\[\lambda \equiv \g{q_{\infty}^2 \over 2 k_F^2}\]
and for $n = 5\times 10^17\,\mrm{cm}^{-3}$, $m = 2m_e$, $\e_F = 1.2\,\mrm{eV}$, $\w_- = 2\e_F/3$, $\eta = 0.2$ and $T = 100\,\mrm{mK}$. Note that here we have take $\eta = 0.2$, which is rather small to allow for a fast convergence and that the coupling strength was tuned manually without tuning any of the other parameters.

As can be seen from Fig.~\ref{phi3d:fig}, the momentum dependance of the order parameter $\phi(i\w,k)$ depends strongly on the coupling strength. At $\lambda = 2$ (most right panel) the order parameter is almost uniform over the entire Fermi sea, while for $\lambda = 0.28$ (most left panel) it is sharply peaked at $k = k_F$. This shows that at weak coupling, when the interaction is not strong enough to excite particles far from the Fermi surface, the pairing is mainly occurring near the Fermi surface. We also note that the structure of $Z(i\w,k)$ and $\chi(i\w,k)$ has a much weaker dependance on the coupling, however their overall amplitude is significantly reduced (see color bars in from Figs.~\ref{Z3d:fig},\ref{chi3d:fig}).

\begin{figure}
\centering
\includegraphics[width=18cm,height=4.3cm]{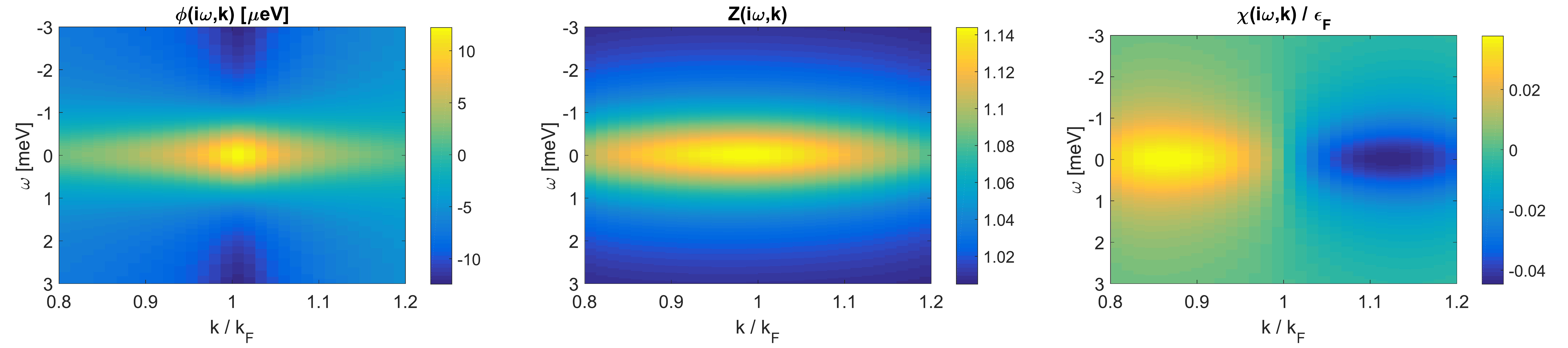}
\caption{The numerical solutions of the Eliashberg equations Eqs.~\ref{phi} restricted to the vicinity of $k = k_F$ for typical parameters used in this paper: $n = 5\times 10^{17}\,\mrm{cm}^{-3}$, $m = 2m_e$, $\e_F = 1.2\,\mrm{eV}$, $\w_- = 2\e_F/3$, $\eta = 0.4$, $\a = 5\times 10^{-16}\mrm{cm}^{3}$ and $T = 20\,\mrm{mK}$.  }\label{fig:phi_r}
\end{figure}

In Fig.~\ref{fig:phi_r} we plot the solutions of Eqs.~5 of the main text restricted to a smaller region near $k = k_F$ with larger $\eta = 0.4$, and with the same parameters except for $T = 20 \, \mrm{mK}$. This solution represents a typical solution for the experimental parameters, and thus represents the self energy for the case of the most dilute superconductor reported in Ref.~\cite{Lin2014}.

The weak coupling behavior of $\phi(i\w,k)$ motivates us to seek a simpler description of the self-energy which is restricted to the Fermi surface. This is obtained by integrating the dispersion in the denominator and the Coulomb interaction analytically over $k'$ in the region $k_F - \d k<k'<k_F +\d k$, where $\d k$ is the cutoff (we use $\d k = k_F /4$ in our calculations, however the results are not very sensitive with respect to the cutoff). We also note that as the coupling is reduced the dispersion renormalization $\chi$ becomes smaller and smaller compared to $\e_F$, and therefore we neglect it.

The resulting {\it isotropic} Eliashberg equations are given by
\begin{align}
&Z(i\w) = 1+ {\lambda T \over\w\, \e_F }\sum_{\w'}{\w' \,Z(i\w' ) \over y(i\w')} { \w_-^2 f_{re} [y(i\w')]\over \w_-^2 +(\w-\w')^2}\label{phi2}\\
&\phi(i\w) = {\lambda T \over \e_F } \sum_{\w'} {\phi(i\w')\over y(i\w')}\left( { \w_-^2 f_{re} [y(i\w')]\over \w_-^2 +(\w-\w')^2} - f_{st} [y(i\w')]\eta\nn \right)
\end{align}
where $y(i\w)\equiv \sqrt{\left[ Z(i\w)\,\w\right]^2+\phi^2(i\w)}/\e_F$ and the functions
\[f_{st}[y] = \int _{-{\d}} ^{\d} dx{y \over y^2 +x^2} \log{2 \over |x|}\]
and
\[f_{re}[y]= \int _{-{\d}} ^{\d} dx {y \over y^2 +x^2} \left(\log{2 \over |x|} -{1\over 2} \log\left[{4+2\lambda_0 \over x^2 +2\lambda_0  }\right]\right)\]
diverge logarithmically at small frequency like $f_{st}[y]\rightarrow\pi {\log{2 \e_F\over y}}$ and $f_{re}[y]\rightarrow\pi {\log{\sqrt{3}\w_- \over 2 y}}$, and go to zero like $1/y$ at large $y$. Here $\lambda_0 \equiv {q_{0}^2 \over 2k_F^2}$.
The logarithmic divergence is the main difference compared to standard Eliashberg theory and is a result of the un-screened Coulomb interactions.

\section{Calculation of the transition temperature}
At $T = T_c$ the solutions of Eq.(\ref{phi2}) go to zero and decouple and therefore, if we are interested only in the transition temperature we can linearize Eq.~(\ref{phi2})
\be
 \phi_\w=  \sum_{\w'} M_{\w,\w'} \phi_{\w'} \label{phi3}
 \ee
where $M_{\w,\w'} = {{\lambda  T_c  }\over \e_F |\w'|}\left( { \w_-^2 f_{re} [\w']\over \w_-^2 +(\w-\w')^2} - f_{st} [\w']\eta\nn \right)$. $T_c$ can be found by seeking when the largest eigenvalue of $M_{\w,\w'} $ becomes unitary. The main advantage of this method is that it involves linear manipulations instead of seeking a solution to the non-linear equation. As a result it is much more stable to large values of $\eta$.

 In Fig.~1 in the main text we plot $T_c$ calculated for two different set of parameters. The blue curves corresponds to $\eta = 0.3$, $m_1 = 4 m_e$, $\a = 8\times 10^{-18}cm^{3}$ and $\w_{sat} = 18\, \mrm{meV}$ and the cyan ones to $\eta = 0.4$, $m_1 = 2 m_e$, $\a = 5.5\times 10^{-16}cm^{3}$ and $\w_{sat} = 11.5\, \mrm{meV}$. We also plot the resulting frequency $\w_-$, the Fermi energy $\e_F$ and the frequency of the transverse mode $\w_T$ in Fig.~1.b and Fig.~1.c of the main text.

\section{The transition temperature in two-dimensions}
In this section we discuss superconductivity in two-dimensional electronic gases based on SrTiO$_3$ (for example, the LaAlO$_3$/SrTiO$_3$, Nb $\delta$-doped SrTiO$_3$ and gated SrTiO$_3$).
\subsection{Paring interaction}
As in 3d, we assume a single resonance model  for the dielectric constant
\be
\e(i\w) \approx \ve_{\infty} +{(\ve_{0}-\ve_{\infty})\w_{T1}^2 \over \w_{T1}^2+\w^2 }
\ee
However, in this case we will assume that the soft mode has completely stiffened and is given by $\w_T = 16\,\mrm{meV}$ [cite Reinle-schmitt], such that $\e_0 \approx 180$.

The 2D polarization bubble of a single band with mass $m$ has the form
\be
\Pi(i\w,q)  =  -{\nu} \left[1-{1\over \sqrt{2}}\sqrt{1-{4 + 4\z^2 \over x^2} +\sqrt{ \left(1-{4 + 4\z^2 \over x^2}\right)^2 +\left({4 \z \over x}\right)^2} } \right]
\ee
where $x \equiv q/k_F$ and $\z \equiv \w k_F / \e_F q  $ and $\nu = m/\pi$.
The RPA interaction is then given by
\be
V_{2d} (i\w,q) ={ q_{\infty}/ \nu\over {q\ve (i\w)/\ve_\infty } - {q_{\infty}   } \Pi(i\w,q)/\nu}\label{V2d}\,.
\ee
where $q_{\infty}  = 2\pi e^2 \nu /\ve_\infty$.
The plasma frequency is now strongly $q$-dependant and is given by
\be
\w_p(q) = \sqrt{ q_{\infty}\; q\over k_F^2} \e_F\,.  \label{wp_2d}
\ee

Just as in the case of three-dimensions, in the limit $\w/\e_F\gg q/k_F$ we can separate the interaction into two resonances
\be
V_{2d}(iw,q) = {q_{\infty} \over \nu q} {1\over \ve(i\w)/\ve_{\infty} + {\w_p(q)^2/\w^2} }  =  {q_{\infty} \over \nu q}\left[ 1-(1-\g) {\w_+ ^2 \over \w_+^2 (q) +\w^2}-\g {\w_- ^2 \over \w_-^2(q) +\w^2} \right]
\ee
\be
\w_\pm ^2(q)  =  {\w_L^2 + \w_p(q)^2 \over 2} \pm \sqrt{\left({\w_L^2 + \w_p(q)^2 \over 2}\right)^2- {\w_p(q) ^2 \w_T^2 }}\,,
\ee
\be
\g \equiv  {\w_T ^2 - \w_-^2 \over  \d\w^2}\,,
\ee
and $\d\w^2\equiv\w^2 _+ - \w^2 _- $.

We may consider two distinct limits. In the limit $\w_p (2k_F) \ll \w_L$ we can simply substitute $\e_\infty/\e_0$ instead of $\g$. On the other hand if $\w_p (2k_F) > \w_L$ than the $q$-dependant plasma frequency crosses through the optical phonon mode as $q$ is integrated from $0$ to roughly $2k_F$, and therefore $\g$ goes to zero.
For the typical Fermi energies in the STO-based 2d gases the latter case holds. As a result the coupling $\g$ will suppress the contribution from the lower the mode $\w_-$ for $q> {2 k_F^2 \over q_{\infty}}{\w_L^2 \over \e_F^2}$.

On the other hand the plasma oscillations appear only in the limit $ {q \over k_F} \ll {\w_p\over \e_F }$ or $q \ll {q_{\infty} \over 2}$.
Comparing these restrictions we find that if $q_{\infty}^2 / k_F^2 \gg \w_L^2 / \e_F^2$ than $\g $ goes to zero much before $q/k_F$ becomes larger than $\w_p(q)/\e_F$.
We therefore argue that the interaction can be approximated by the plasmon pole approximation
\be
V_{2d}(i\w,q) = {q_{\infty}\g(q) \over \nu q}\left[ 1- {\w_-(q) ^2 \over \w_-^2(q) +\w^2} \right]\label{V2d_app}
\ee

\subsection{Linearized Eliashberg equations}
Just as in 3d we linearize the Eliashberg equations
\begin{align}
&\phi(i\w) = {1\over \b_c\,N} \sum_{\w'}\int_{k_F - \d k_F} ^{k_F + \d k_F} {d k\over 2\pi} \int_{0}^{2\pi} {d\t \over 2\pi}\; k' {\phi(i\w')\over \w'^2 + \e_{k'}^2}\left[V_{re}(i\w-i\w', |\bs k_F -\bs k'|) - \eta V_{st}( |\bs k_F -\bs k'|) \right]
\end{align}
where
\[V_{st}( q) = { \g(q)q_{\infty} \over \nu q}\]
and
\[V_{re}(i\w, q)\equiv {1\over \nu}\left[ {q_{\infty}\g(q) \over  q } -{q_0 \over q + q_0}  \right]{\w_-(q) ^2 \over \w_-^2(q) +\w^2}\]
where we have restricted the integration close to the Fermi momentum $\d k_F \ll k_F$ (assuming that the coupling is weak we take $\d k_F = k_F / 8$) and we taken into account the finite value of the interaction at $\w \rightarrow 0$ as in Fig.~\ref{fig:V}. Note that here the average over the angle of $\bs k'$ is performed numerically because both $\w_p$ and $\g$ are funcitons of $q = |\bs k- \bs k'|$.

As before we have the eigenvalue problem
\be
 \sum_{\w'} M_{\w,\w'}\phi_{\w'} = \phi_\w
\ee
where the matrix $K$ is given by
\be
M_{\w,\w'} =  {1\over \b_c} \int_{k_F-\delta k_F}^{k_F + \delta k_F} { dk' \over 2\pi} {d\t \over 2\pi}\,{k' \over \w'^2 + \e_{k'}^2}\left[V_{re}(i\w-i\w', |\bs k_F -\bs k'|) - \eta V_{st}(|\bs k_F -\bs k'|) \right] \label{K}
\ee
where the Matsubara frequencies $\w$ and $\w'$ are spaced by $2\pi /\b_c$ and run up to some cutoff.
$T_c$ is obtained when $M $ has an eigenvalue of unity. In fact, $T_c$ corresponds to the temperature where the largest eigenvalue of $M$ becomes unity.

The resulting $T_c$ for $m = m_e$, $\w_T = 16 \mrm{meV}$ and various $\eta $'s is plotted in Fig.~\ref{fig:2d}.
In Fig.~\ref{fig:WF_2d} we also plot the eigenvector $\phi_{\w}/\phi_0$ at $T_c$ with $\w_T = 16\,\mrm{meV}$ and $\eta = 0.25$ for three different densities corresponding to Fermi energies of $\e_F = 36, 48 \,\mrm{and}\,60\,\mrm{meV}$. The dashed line is a Lorentzian shape with width $\w_T$ for comparison. As can be seen, the width of the $\phi_\w$, which mimics the width of the retarded interaction is approximately $4\,\mrm{meV}$. Therefore it is significantly smaller than $\e_F$ (and also $\w_T$). This justifies the use of a small $\eta$ at the typical range of densities where superconductivity is observed.
Indeed we expect the width of the interaction in frequency space to be significantly smaller than $\w_p(q = 2k_F)$ because most of the weight in the angular integral comes from small $q$ where $\w_p(q) \ll \e_F ,\w_T$.
We also note that the shape is not exactly a Lorentzian (namely, it has long tails).

\begin{figure}
\centering
\includegraphics[width=8cm,height=7cm]{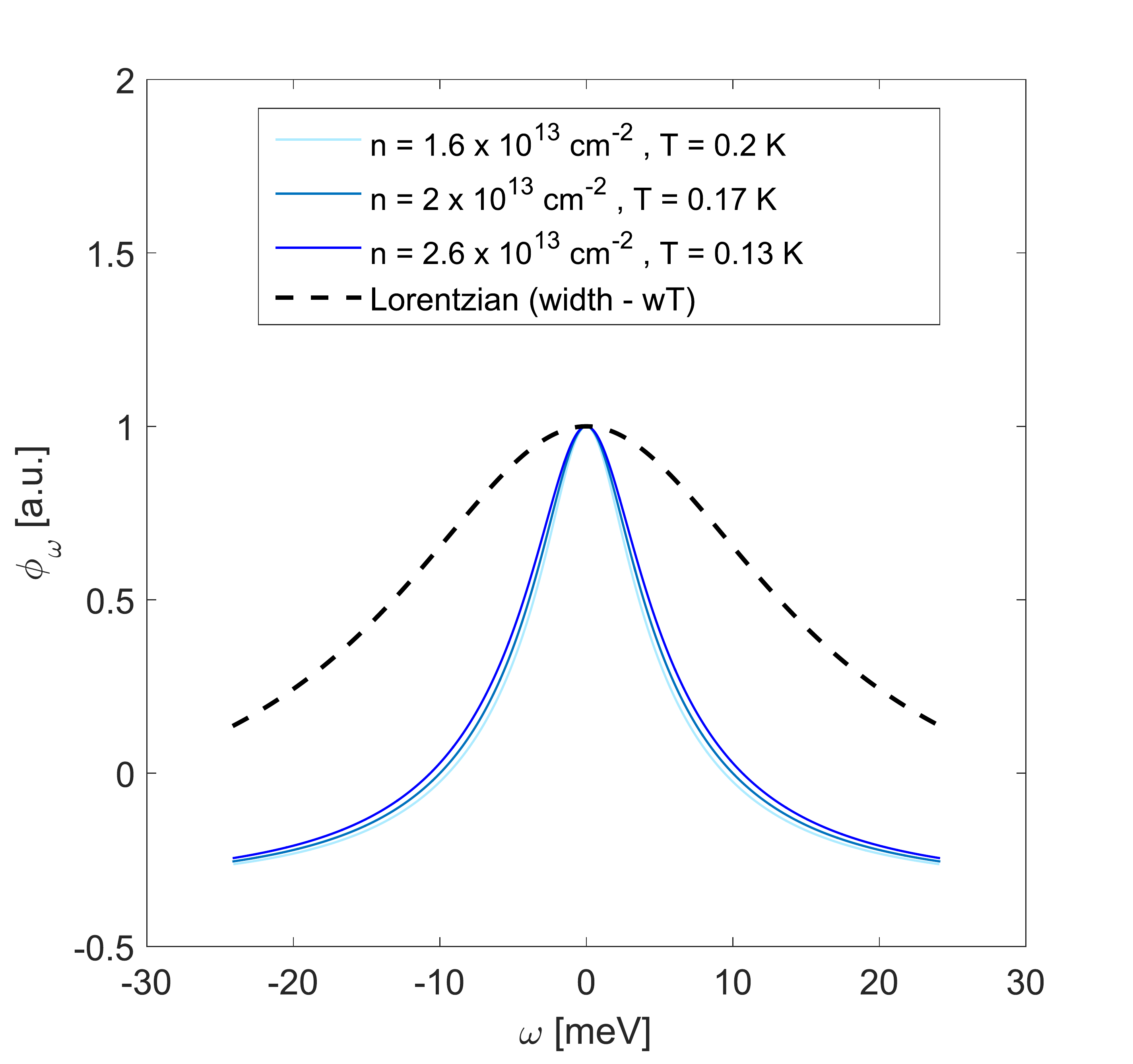}
\caption{the eigenvector $\phi_{\w}/\phi_0$ at $T_c$ with $\w_T = 16\,\mrm{meV}$ and $\eta = 0.25$ for three different densities corresponding to Fermi energies of $\e_F = 36, 48 \,\mrm{and}\,60\,\mrm{meV}$. This figure shows that the width of the angle averaged retarded interaction is significantly smaller than $\e_F$. }\label{fig:WF_2d}
\end{figure}

\section{Analytic continuation of the self-energy}
In this section we elaborate on the controlled Pad\'e approximation~\cite{Beach2000} used to analytically continue the self-energy in the Green's function (Eq.~6 in the main text) to the real axis.
From Eq.~\ref{phi2} we obtain the functions $\phi(i\w)$ and $Z(i\w)$ in a finite number of fermionic Matsubara frequencies lying in the region $|\w_n|<\Omega$. There is no unique analytic continuation of a finite set of points to the entire upper half plane. The Pad\'e form
\be
\Sigma(z) = {P_{r-1}(z) \over Q_{r}(z)}\,,\label{self_energy}
\ee
where
\[P_{r-1}(z) = \sum_{l = 0}^{r-1} p_l z^l\]
and
\[Q_r(z)\sum_{l = 0}^{r} q_l z^l\,,\]
is often used because it posses all the analytic properties of a response function in the upper half plane. This statement is actually true under the condition that $p_{r-1}$ is real and positive. Therefore, Ref.~\cite{Beach2000} has proposed to use the imaginary part of $p_{r-1}$ as a control parameter to quantify the quality of the analytic continuation.
Following Ref.~\cite{Beach2000}, we analytically continue from $r$ Matsubara points which, i.e. $\left\{ i\w_j \right\}_{j=1}^{r}$ (note that $r$ need not be the full number of Matsubara frequencies for which $\phi(i\w)$ and $Z(i\w)$ is known). The coefficients of these polynomials are obtained by solving a linear set of equations
\[ \sum_{n = 0}^{r-1 }\left(i\w_j \right)^n\,p_r \, -\Sigma(i\w_j)\sum_{n = 0}^{r-1 }\left(i\w_j \right)^n\,q_r = \Sigma(i\w_j)\,(i\w_j)^{r}  \]
for the set of Matsubara points $z\in \left\{ i\w_j \right\}_{j=1}^{r}$.
The imaginary part of $p_{r-1}$ is monitored and found to be smaller than numerical precision in all calculations.

\begin{figure}
\centering
\includegraphics[width=8cm,height=7cm]{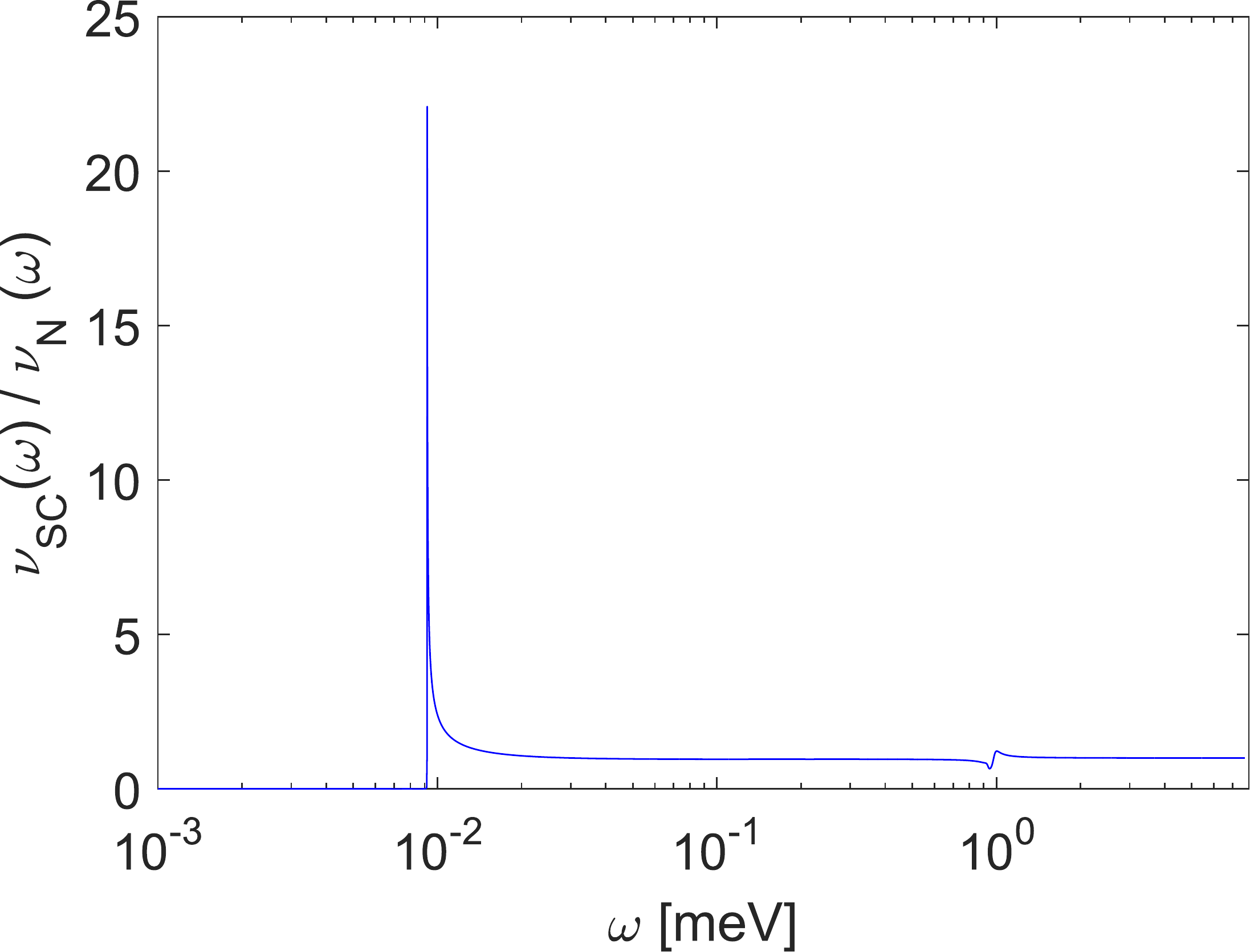}
\caption{The TDOS calculated from Eq.~\ref{TDOS} for $n = 5\times 10^{17}\,\mrm{cm}^{-3}$ and the same parameters used for the cyan curve in Fig.~\ref{fig:domes} and Fig.~\ref{fig:TDOS}. }\label{fig:WF_2d}
\end{figure}

\section{the tunneling density of states in a dilute superconductor }
Given the analytic continuation of the of the self-energy Eq.~\ref{self_energy} we can now calculate the single-particle density of states from the imaginary part of the {\it electron}'s Green's function~\cite{Schrieffer1963}
\be
G_e (\w,k) = { \w + \e_k / Z(\w) \over \w^2  - \e_k^2 /Z^2(\w) - \D^2(\w)+i 0^+ }
\ee
In the case of a shallow band ($\w \sim \e_F$) this gives
\begin{align}
\nu_{sc}(\w) = \int{d^3 k \over (2\pi)^3} \,\mrm{Im}\,G(\w+ i 0^+,k) \label{TDOS} = \nu\, \mrm{Re} \left[ \sqrt{1-{Z(\w)E_\w/ \e_F}}\left|  \w -E_{ \w}\right|+ \sqrt{1+{Z(\w)E_\w/ \e_F}}\left|  \w+E_{\w}\right| \over 2\, E_{ \w}\right]
\end{align}
where $\nu$ is the density of states at the Fermi level, $E_{ \w} \equiv\sqrt{ \w ^2-\D^2(\w)}$, $\D(i\w) = \phi(i\w)/Z(i\w)$ and we have neglected the momentum dependence of the gap and the density of states coming from the smaller Fermi pockets.

\begin{thebibliography}{52}%
\makeatletter
\providecommand \@ifxundefined [1]{%
 \@ifx{#1\undefined}
}%
\providecommand \@ifnum [1]{%
 \ifnum #1\expandafter \@firstoftwo
 \else \expandafter \@secondoftwo
 \fi
}%
\providecommand \@ifx [1]{%
 \ifx #1\expandafter \@firstoftwo
 \else \expandafter \@secondoftwo
 \fi
}%
\providecommand \natexlab [1]{#1}%
\providecommand \enquote  [1]{``#1''}%
\providecommand \bibnamefont  [1]{#1}%
\providecommand \bibfnamefont [1]{#1}%
\providecommand \citenamefont [1]{#1}%
\providecommand \href@noop [0]{\@secondoftwo}%
\providecommand \href [0]{\begingroup \@sanitize@url \@href}%
\providecommand \@href[1]{\@@startlink{#1}\@@href}%
\providecommand \@@href[1]{\endgroup#1\@@endlink}%
\providecommand \@sanitize@url [0]{\catcode `\\12\catcode `\$12\catcode
  `\&12\catcode `\#12\catcode `\^12\catcode `\_12\catcode `\%12\relax}%
\providecommand \@@startlink[1]{}%
\providecommand \@@endlink[0]{}%
\providecommand \url  [0]{\begingroup\@sanitize@url \@url }%
\providecommand \@url [1]{\endgroup\@href {#1}{\urlprefix }}%
\providecommand \urlprefix  [0]{URL }%
\providecommand \Eprint [0]{\href }%
\providecommand \doibase [0]{http://dx.doi.org/}%
\providecommand \selectlanguage [0]{\@gobble}%
\providecommand \bibinfo  [0]{\@secondoftwo}%
\providecommand \bibfield  [0]{\@secondoftwo}%
\providecommand \translation [1]{[#1]}%
\providecommand \BibitemOpen [0]{}%
\providecommand \bibitemStop [0]{}%
\providecommand \bibitemNoStop [0]{.\EOS\space}%
\providecommand \EOS [0]{\spacefactor3000\relax}%
\providecommand \BibitemShut  [1]{\csname bibitem#1\endcsname}%
\let\auto@bib@innerbib\@empty
%</preamble>
\bibitem [{\citenamefont {Bogoliubov}\ \emph {et~al.}(1960)\citenamefont
  {Bogoliubov}, \citenamefont {Tolmachev},\ and\ \citenamefont
  {Shirkov}}]{bogoliubov1960new}%
  \BibitemOpen
  \bibfield  {author} {\bibinfo {author} {\bibfnamefont {N.~N.}\ \bibnamefont
  {Bogoliubov}}, \bibinfo {author} {\bibfnamefont {V.}~\bibnamefont
  {Tolmachev}}, \ and\ \bibinfo {author} {\bibfnamefont {D.}~\bibnamefont
  {Shirkov}},\ }\href@noop {} {\bibfield  {journal} {\bibinfo  {journal}
  {Consultants Bureau, New York}\ } (\bibinfo {year} {1960})}\BibitemShut
  {NoStop}%
\bibitem [{\citenamefont {Schooley}\ \emph {et~al.}(1964)\citenamefont
  {Schooley}, \citenamefont {Hosler},\ and\ \citenamefont
  {Cohen}}]{Schooley1964}%
  \BibitemOpen
  \bibfield  {author} {\bibinfo {author} {\bibfnamefont {J.}~\bibnamefont
  {Schooley}}, \bibinfo {author} {\bibfnamefont {W.}~\bibnamefont {Hosler}}, \
  and\ \bibinfo {author} {\bibfnamefont {M.~L.}\ \bibnamefont {Cohen}},\
  }\href@noop {} {\bibfield  {journal} {\bibinfo  {journal} {Physical Review
  Letters}\ }\textbf {\bibinfo {volume} {12}},\ \bibinfo {pages} {474}
  (\bibinfo {year} {1964})}\BibitemShut {NoStop}%
\bibitem [{\citenamefont {Koonce}\ \emph {et~al.}(1967)\citenamefont {Koonce},
  \citenamefont {Cohen}, \citenamefont {Schooley}, \citenamefont {Hosler},\
  and\ \citenamefont {Pfeiffer}}]{Koonce1967}%
  \BibitemOpen
  \bibfield  {author} {\bibinfo {author} {\bibfnamefont {C.}~\bibnamefont
  {Koonce}}, \bibinfo {author} {\bibfnamefont {M.~L.}\ \bibnamefont {Cohen}},
  \bibinfo {author} {\bibfnamefont {J.}~\bibnamefont {Schooley}}, \bibinfo
  {author} {\bibfnamefont {W.}~\bibnamefont {Hosler}}, \ and\ \bibinfo {author}
  {\bibfnamefont {E.}~\bibnamefont {Pfeiffer}},\ }\href@noop {} {\bibfield
  {journal} {\bibinfo  {journal} {Physical Review}\ }\textbf {\bibinfo {volume}
  {163}},\ \bibinfo {pages} {380} (\bibinfo {year} {1967})}\BibitemShut
  {NoStop}%
\bibitem [{\citenamefont {Binnig}\ \emph {et~al.}(1980)\citenamefont {Binnig},
  \citenamefont {Baratoff}, \citenamefont {Hoenig},\ and\ \citenamefont
  {Bednorz}}]{Binnig1980}%
  \BibitemOpen
  \bibfield  {author} {\bibinfo {author} {\bibfnamefont {G.}~\bibnamefont
  {Binnig}}, \bibinfo {author} {\bibfnamefont {A.}~\bibnamefont {Baratoff}},
  \bibinfo {author} {\bibfnamefont {H.~E.}\ \bibnamefont {Hoenig}}, \ and\
  \bibinfo {author} {\bibfnamefont {J.~G.}\ \bibnamefont {Bednorz}},\ }\href
  {\doibase 10.1103/PhysRevLett.45.1352} {\bibfield  {journal} {\bibinfo
  {journal} {Phys. Rev. Lett.}\ }\textbf {\bibinfo {volume} {45}},\ \bibinfo
  {pages} {1352} (\bibinfo {year} {1980})}\BibitemShut {NoStop}%
\bibitem [{\citenamefont {Baratoff}\ and\ \citenamefont
  {Binnig}(1981)}]{Binnig1981}%
  \BibitemOpen
  \bibfield  {author} {\bibinfo {author} {\bibfnamefont {A.}~\bibnamefont
  {Baratoff}}\ and\ \bibinfo {author} {\bibfnamefont {G.}~\bibnamefont
  {Binnig}},\ }\href@noop {} {\bibfield  {journal} {\bibinfo  {journal}
  {Physica B+ C}\ }\textbf {\bibinfo {volume} {108}},\ \bibinfo {pages} {1335}
  (\bibinfo {year} {1981})}\BibitemShut {NoStop}%
\bibitem [{\citenamefont {Lin}\ \emph {et~al.}(2014)\citenamefont {Lin},
  \citenamefont {Bridoux}, \citenamefont {Gourgout}, \citenamefont {Seyfarth},
  \citenamefont {Kr{\"{a}}mer}, \citenamefont {Nardone}, \citenamefont
  {Fauqu{\'{e}}},\ and\ \citenamefont {Behnia}}]{Lin20133}%
  \BibitemOpen
  \bibfield  {author} {\bibinfo {author} {\bibfnamefont {X.}~\bibnamefont
  {Lin}}, \bibinfo {author} {\bibfnamefont {G.}~\bibnamefont {Bridoux}},
  \bibinfo {author} {\bibfnamefont {A.}~\bibnamefont {Gourgout}}, \bibinfo
  {author} {\bibfnamefont {G.}~\bibnamefont {Seyfarth}}, \bibinfo {author}
  {\bibfnamefont {S.}~\bibnamefont {Kr{\"{a}}mer}}, \bibinfo {author}
  {\bibfnamefont {M.}~\bibnamefont {Nardone}}, \bibinfo {author} {\bibfnamefont
  {B.}~\bibnamefont {Fauqu{\'{e}}}}, \ and\ \bibinfo {author} {\bibfnamefont
  {K.}~\bibnamefont {Behnia}},\ }\href {\doibase
  10.1103/PhysRevLett.112.207002} {\bibfield  {journal} {\bibinfo  {journal}
  {Physical Review Letters}\ }\textbf {\bibinfo {volume} {112}},\ \bibinfo
  {pages} {207002} (\bibinfo {year} {2014})}\BibitemShut {NoStop}%
\bibitem [{\citenamefont {Lin}\ \emph {et~al.}(2013)\citenamefont {Lin},
  \citenamefont {Zhu}, \citenamefont {Fauqu{\'{e}}},\ and\ \citenamefont
  {Behnia}}]{Lin2014}%
  \BibitemOpen
  \bibfield  {author} {\bibinfo {author} {\bibfnamefont {X.}~\bibnamefont
  {Lin}}, \bibinfo {author} {\bibfnamefont {Z.}~\bibnamefont {Zhu}}, \bibinfo
  {author} {\bibfnamefont {B.}~\bibnamefont {Fauqu{\'{e}}}}, \ and\ \bibinfo
  {author} {\bibfnamefont {K.}~\bibnamefont {Behnia}},\ }\href {\doibase
  10.1103/PhysRevX.3.021002} {\bibfield  {journal} {\bibinfo  {journal}
  {Physical Review X}\ }\textbf {\bibinfo {volume} {3}},\ \bibinfo {pages}
  {021002} (\bibinfo {year} {2013})}\BibitemShut {NoStop}%
\bibitem [{\citenamefont {Weaver}(1959)}]{weaver1959}%
  \BibitemOpen
  \bibfield  {author} {\bibinfo {author} {\bibfnamefont {H.}~\bibnamefont
  {Weaver}},\ }\href@noop {} {\bibfield  {journal} {\bibinfo  {journal}
  {Journal of Physics and Chemistry of Solids}\ }\textbf {\bibinfo {volume}
  {11}},\ \bibinfo {pages} {274} (\bibinfo {year} {1959})}\BibitemShut
  {NoStop}%
\bibitem [{\citenamefont {M\"uller}\ and\ \citenamefont
  {Burkard}(1979)}]{Muller1979}%
  \BibitemOpen
  \bibfield  {author} {\bibinfo {author} {\bibfnamefont {K.~A.}\ \bibnamefont
  {M\"uller}}\ and\ \bibinfo {author} {\bibfnamefont {H.}~\bibnamefont
  {Burkard}},\ }\href {\doibase 10.1103/PhysRevB.19.3593} {\bibfield  {journal}
  {\bibinfo  {journal} {Phys. Rev. B}\ }\textbf {\bibinfo {volume} {19}},\
  \bibinfo {pages} {3593} (\bibinfo {year} {1979})}\BibitemShut {NoStop}%
\bibitem [{\citenamefont {Vogt}(1995)}]{Vogt1995}%
  \BibitemOpen
  \bibfield  {author} {\bibinfo {author} {\bibfnamefont {H.}~\bibnamefont
  {Vogt}},\ }\href {\doibase 10.1103/PhysRevB.51.8046} {\bibfield  {journal}
  {\bibinfo  {journal} {Phys. Rev. B}\ }\textbf {\bibinfo {volume} {51}},\
  \bibinfo {pages} {8046} (\bibinfo {year} {1995})}\BibitemShut {NoStop}%
\bibitem [{\citenamefont {Vogt}\ and\ \citenamefont
  {Rossbroich}(1981)}]{Vogt1981}%
  \BibitemOpen
  \bibfield  {author} {\bibinfo {author} {\bibfnamefont {H.}~\bibnamefont
  {Vogt}}\ and\ \bibinfo {author} {\bibfnamefont {G.}~\bibnamefont
  {Rossbroich}},\ }\href {\doibase 10.1103/PhysRevB.24.3086} {\bibfield
  {journal} {\bibinfo  {journal} {Phys. Rev. B}\ }\textbf {\bibinfo {volume}
  {24}},\ \bibinfo {pages} {3086} (\bibinfo {year} {1981})}\BibitemShut
  {NoStop}%
\bibitem [{\citenamefont {Kamarás}\ \emph {et~al.}(1995)\citenamefont
  {Kamarás}, \citenamefont {Barth}, \citenamefont {Keilmann}, \citenamefont
  {Henn}, \citenamefont {Reedyk}, \citenamefont {Thomsen}, \citenamefont
  {Cardona}, \citenamefont {Kircher}, \citenamefont {Richards},\ and\
  \citenamefont {Stehlé}}]{Kamaras1995}%
  \BibitemOpen
  \bibfield  {author} {\bibinfo {author} {\bibfnamefont {K.}~\bibnamefont
  {Kamarás}}, \bibinfo {author} {\bibfnamefont {K.-L.}\ \bibnamefont
  {Barth}}, \bibinfo {author} {\bibfnamefont {F.}~\bibnamefont {Keilmann}},
  \bibinfo {author} {\bibfnamefont {R.}~\bibnamefont {Henn}}, \bibinfo {author}
  {\bibfnamefont {M.}~\bibnamefont {Reedyk}}, \bibinfo {author} {\bibfnamefont
  {C.}~\bibnamefont {Thomsen}}, \bibinfo {author} {\bibfnamefont
  {M.}~\bibnamefont {Cardona}}, \bibinfo {author} {\bibfnamefont
  {J.}~\bibnamefont {Kircher}}, \bibinfo {author} {\bibfnamefont {P.~L.}\
  \bibnamefont {Richards}}, \ and\ \bibinfo {author} {\bibfnamefont {J.-L.}\
  \bibnamefont {Stehlé}},\ }\href {\doibase 10.1063/1.360364} {\bibfield
  {journal} {\bibinfo  {journal} {Journal of Applied Physics}\ }\textbf
  {\bibinfo {volume} {78}},\ \bibinfo {pages} {1235} (\bibinfo {year}
  {1995})}\BibitemShut {NoStop}%
\bibitem [{\citenamefont {van Mechelen}\ \emph {et~al.}(2008)\citenamefont {van
  Mechelen}, \citenamefont {van~der Marel}, \citenamefont {Grimaldi},
  \citenamefont {Kuzmenko}, \citenamefont {Armitage}, \citenamefont {Reyren},
  \citenamefont {Hagemann},\ and\ \citenamefont {Mazin}}]{vanderMarel2008}%
  \BibitemOpen
  \bibfield  {author} {\bibinfo {author} {\bibfnamefont {J.~L.~M.}\
  \bibnamefont {van Mechelen}}, \bibinfo {author} {\bibfnamefont
  {D.}~\bibnamefont {van~der Marel}}, \bibinfo {author} {\bibfnamefont
  {C.}~\bibnamefont {Grimaldi}}, \bibinfo {author} {\bibfnamefont {A.~B.}\
  \bibnamefont {Kuzmenko}}, \bibinfo {author} {\bibfnamefont {N.~P.}\
  \bibnamefont {Armitage}}, \bibinfo {author} {\bibfnamefont {N.}~\bibnamefont
  {Reyren}}, \bibinfo {author} {\bibfnamefont {H.}~\bibnamefont {Hagemann}}, \
  and\ \bibinfo {author} {\bibfnamefont {I.~I.}\ \bibnamefont {Mazin}},\ }\href
  {\doibase 10.1103/PhysRevLett.100.226403} {\bibfield  {journal} {\bibinfo
  {journal} {Phys. Rev. Lett.}\ }\textbf {\bibinfo {volume} {100}},\ \bibinfo
  {pages} {226403} (\bibinfo {year} {2008})}\BibitemShut {NoStop}%
\bibitem [{\citenamefont {Gurevich}\ \emph {et~al.}(1962)\citenamefont
  {Gurevich}, \citenamefont {Larkin},\ and\ \citenamefont
  {Firsov}}]{Gurevich1962}%
  \BibitemOpen
  \bibfield  {author} {\bibinfo {author} {\bibfnamefont {L.~V.}\ \bibnamefont
  {Gurevich}}, \bibinfo {author} {\bibfnamefont {A.~I.}\ \bibnamefont
  {Larkin}}, \ and\ \bibinfo {author} {\bibfnamefont {Y.~A.}\ \bibnamefont
  {Firsov}},\ }\href@noop {} {\bibfield  {journal} {\bibinfo  {journal} {Sov.
  Phys. Sol. State}\ }\textbf {\bibinfo {volume} {4}},\ \bibinfo {pages} {131}
  (\bibinfo {year} {1962})}\BibitemShut {NoStop}%
\bibitem [{\citenamefont {Takada}(1980)}]{Takada1980}%
  \BibitemOpen
  \bibfield  {author} {\bibinfo {author} {\bibfnamefont {Y.}~\bibnamefont
  {Takada}},\ }\href {\doibase 10.1143/JPSJ.49.1267} {\bibfield  {journal}
  {\bibinfo  {journal} {Journal of the Physical Society of Japan}\ }\textbf
  {\bibinfo {volume} {49}},\ \bibinfo {pages} {1267} (\bibinfo {year}
  {1980})}\BibitemShut {NoStop}%
\bibitem [{\citenamefont {Kirzhnits}\ \emph {et~al.}(1973)\citenamefont
  {Kirzhnits}, \citenamefont {Maksimov},\ and\ \citenamefont
  {Khomskii}}]{kirzhnits1973description}%
  \BibitemOpen
  \bibfield  {author} {\bibinfo {author} {\bibfnamefont {D.}~\bibnamefont
  {Kirzhnits}}, \bibinfo {author} {\bibfnamefont {E.}~\bibnamefont {Maksimov}},
  \ and\ \bibinfo {author} {\bibfnamefont {D.}~\bibnamefont {Khomskii}},\
  }\href@noop {} {\bibfield  {journal} {\bibinfo  {journal} {Journal of Low
  Temperature Physics}\ }\textbf {\bibinfo {volume} {10}},\ \bibinfo {pages}
  {79} (\bibinfo {year} {1973})}\BibitemShut {NoStop}%
\bibitem [{\citenamefont {Grabowski}\ and\ \citenamefont
  {Sham}(1984)}]{Grabowski1984}%
  \BibitemOpen
  \bibfield  {author} {\bibinfo {author} {\bibfnamefont {M.}~\bibnamefont
  {Grabowski}}\ and\ \bibinfo {author} {\bibfnamefont {L.~J.}\ \bibnamefont
  {Sham}},\ }\href {\doibase 10.1103/PhysRevB.29.6132} {\bibfield  {journal}
  {\bibinfo  {journal} {Physical Review B}\ }\textbf {\bibinfo {volume} {29}},\
  \bibinfo {pages} {6132} (\bibinfo {year} {1984})}\BibitemShut {NoStop}%
\bibitem [{\citenamefont {Grimaldi}\ \emph {et~al.}(1995)\citenamefont
  {Grimaldi}, \citenamefont {Pietronero},\ and\ \citenamefont
  {Str\"assler}}]{Grimaldi1995}%
  \BibitemOpen
  \bibfield  {author} {\bibinfo {author} {\bibfnamefont {C.}~\bibnamefont
  {Grimaldi}}, \bibinfo {author} {\bibfnamefont {L.}~\bibnamefont
  {Pietronero}}, \ and\ \bibinfo {author} {\bibfnamefont {S.}~\bibnamefont
  {Str\"assler}},\ }\href {\doibase 10.1103/PhysRevLett.75.1158} {\bibfield
  {journal} {\bibinfo  {journal} {Phys. Rev. Lett.}\ }\textbf {\bibinfo
  {volume} {75}},\ \bibinfo {pages} {1158} (\bibinfo {year}
  {1995})}\BibitemShut {NoStop}%
\bibitem [{\citenamefont {Klimin}\ \emph {et~al.}(2014)\citenamefont {Klimin},
  \citenamefont {Tempere}, \citenamefont {Devreese},\ and\ \citenamefont
  {van~der Marel}}]{Klimin2014}%
  \BibitemOpen
  \bibfield  {author} {\bibinfo {author} {\bibfnamefont {S.~N.}\ \bibnamefont
  {Klimin}}, \bibinfo {author} {\bibfnamefont {J.}~\bibnamefont {Tempere}},
  \bibinfo {author} {\bibfnamefont {J.~T.}\ \bibnamefont {Devreese}}, \ and\
  \bibinfo {author} {\bibfnamefont {D.}~\bibnamefont {van~der Marel}},\ }\href
  {\doibase 10.1103/PhysRevB.89.184514} {\bibfield  {journal} {\bibinfo
  {journal} {Phys. Rev. B}\ }\textbf {\bibinfo {volume} {89}},\ \bibinfo
  {pages} {184514} (\bibinfo {year} {2014})}\BibitemShut {NoStop}%
\bibitem [{\citenamefont {Gor`kov}(2016)}]{gor2015phonon}%
  \BibitemOpen
  \bibfield  {author} {\bibinfo {author} {\bibfnamefont {L.~P.}\ \bibnamefont
  {Gor`kov}},\ }\href {\doibase 10.1073/pnas.1604145113} {\bibfield  {journal}
  {\bibinfo  {journal} {Proc. Natl. Acad. Sci.}\ }\textbf {\bibinfo {volume}
  {113}},\ \bibinfo {pages} {4646} (\bibinfo {year} {2016})}\BibitemShut
  {NoStop}%
\bibitem [{\citenamefont {Edge}\ \emph {et~al.}(2015)\citenamefont {Edge},
  \citenamefont {Kedem}, \citenamefont {Aschauer}, \citenamefont {Spaldin},\
  and\ \citenamefont {Balatsky}}]{Edge2015}%
  \BibitemOpen
  \bibfield  {author} {\bibinfo {author} {\bibfnamefont {J.~M.}\ \bibnamefont
  {Edge}}, \bibinfo {author} {\bibfnamefont {Y.}~\bibnamefont {Kedem}},
  \bibinfo {author} {\bibfnamefont {U.}~\bibnamefont {Aschauer}}, \bibinfo
  {author} {\bibfnamefont {N.~A.}\ \bibnamefont {Spaldin}}, \ and\ \bibinfo
  {author} {\bibfnamefont {A.~V.}\ \bibnamefont {Balatsky}},\ }\href {\doibase
  10.1103/PhysRevLett.115.247002} {\bibfield  {journal} {\bibinfo  {journal}
  {Phys. Rev. Lett.}\ }\textbf {\bibinfo {volume} {115}},\ \bibinfo {pages}
  {247002} (\bibinfo {year} {2015})}\BibitemShut {NoStop}%
\bibitem [{\citenamefont {B{\"a}uerle}\ \emph {et~al.}(1980)\citenamefont
  {B{\"a}uerle}, \citenamefont {Wagner}, \citenamefont {W{\"o}hlecke},
  \citenamefont {Dorner},\ and\ \citenamefont
  {Kraxenberger}}]{bauerle1980soft}%
  \BibitemOpen
  \bibfield  {author} {\bibinfo {author} {\bibfnamefont {D.}~\bibnamefont
  {B{\"a}uerle}}, \bibinfo {author} {\bibfnamefont {D.}~\bibnamefont {Wagner}},
  \bibinfo {author} {\bibfnamefont {M.}~\bibnamefont {W{\"o}hlecke}}, \bibinfo
  {author} {\bibfnamefont {B.}~\bibnamefont {Dorner}}, \ and\ \bibinfo {author}
  {\bibfnamefont {H.}~\bibnamefont {Kraxenberger}},\ }\href@noop {} {\bibfield
  {journal} {\bibinfo  {journal} {Zeitschrift f{\"u}r Physik B Condensed
  Matter}\ }\textbf {\bibinfo {volume} {38}},\ \bibinfo {pages} {335} (\bibinfo
  {year} {1980})}\BibitemShut {NoStop}%
\bibitem [{\citenamefont {Crandles}\ \emph {et~al.}(1999)\citenamefont
  {Crandles}, \citenamefont {Nicholas}, \citenamefont {Dreher}, \citenamefont
  {Homes}, \citenamefont {McConnell}, \citenamefont {Clayman}, \citenamefont
  {Gong},\ and\ \citenamefont {Greedan}}]{Crandles1999}%
  \BibitemOpen
  \bibfield  {author} {\bibinfo {author} {\bibfnamefont {D.~A.}\ \bibnamefont
  {Crandles}}, \bibinfo {author} {\bibfnamefont {B.}~\bibnamefont {Nicholas}},
  \bibinfo {author} {\bibfnamefont {C.}~\bibnamefont {Dreher}}, \bibinfo
  {author} {\bibfnamefont {C.~C.}\ \bibnamefont {Homes}}, \bibinfo {author}
  {\bibfnamefont {A.~W.}\ \bibnamefont {McConnell}}, \bibinfo {author}
  {\bibfnamefont {B.~P.}\ \bibnamefont {Clayman}}, \bibinfo {author}
  {\bibfnamefont {W.~H.}\ \bibnamefont {Gong}}, \ and\ \bibinfo {author}
  {\bibfnamefont {J.~E.}\ \bibnamefont {Greedan}},\ }\href {\doibase
  10.1103/PhysRevB.59.12842} {\bibfield  {journal} {\bibinfo  {journal} {Phys.
  Rev. B}\ }\textbf {\bibinfo {volume} {59}},\ \bibinfo {pages} {12842}
  (\bibinfo {year} {1999})}\BibitemShut {NoStop}%
\bibitem [{\citenamefont {Caviglia}\ \emph {et~al.}(2008)\citenamefont
  {Caviglia}, \citenamefont {Gariglio}, \citenamefont {Reyren}, \citenamefont
  {Jaccard}, \citenamefont {Schneider}, \citenamefont {Gabay}, \citenamefont
  {Thiel}, \citenamefont {Hammerl}, \citenamefont {Mannhart},\ and\
  \citenamefont {Triscone}}]{caviglia2008electric}%
  \BibitemOpen
  \bibfield  {author} {\bibinfo {author} {\bibfnamefont {A.}~\bibnamefont
  {Caviglia}}, \bibinfo {author} {\bibfnamefont {S.}~\bibnamefont {Gariglio}},
  \bibinfo {author} {\bibfnamefont {N.}~\bibnamefont {Reyren}}, \bibinfo
  {author} {\bibfnamefont {D.}~\bibnamefont {Jaccard}}, \bibinfo {author}
  {\bibfnamefont {T.}~\bibnamefont {Schneider}}, \bibinfo {author}
  {\bibfnamefont {M.}~\bibnamefont {Gabay}}, \bibinfo {author} {\bibfnamefont
  {S.}~\bibnamefont {Thiel}}, \bibinfo {author} {\bibfnamefont
  {G.}~\bibnamefont {Hammerl}}, \bibinfo {author} {\bibfnamefont
  {J.}~\bibnamefont {Mannhart}}, \ and\ \bibinfo {author} {\bibfnamefont
  {J.-M.}\ \bibnamefont {Triscone}},\ }\href@noop {} {\bibfield  {journal}
  {\bibinfo  {journal} {Nature}\ }\textbf {\bibinfo {volume} {456}},\ \bibinfo
  {pages} {624} (\bibinfo {year} {2008})}\BibitemShut {NoStop}%
\bibitem [{\citenamefont {Ueno}\ \emph {et~al.}(2008)\citenamefont {Ueno},
  \citenamefont {Nakamura}, \citenamefont {Shimotani}, \citenamefont {Ohtomo},
  \citenamefont {Kimura}, \citenamefont {Nojima}, \citenamefont {Aoki},
  \citenamefont {Iwasa},\ and\ \citenamefont {Kawasaki}}]{ueno2008electric}%
  \BibitemOpen
  \bibfield  {author} {\bibinfo {author} {\bibfnamefont {K.}~\bibnamefont
  {Ueno}}, \bibinfo {author} {\bibfnamefont {S.}~\bibnamefont {Nakamura}},
  \bibinfo {author} {\bibfnamefont {H.}~\bibnamefont {Shimotani}}, \bibinfo
  {author} {\bibfnamefont {A.}~\bibnamefont {Ohtomo}}, \bibinfo {author}
  {\bibfnamefont {N.}~\bibnamefont {Kimura}}, \bibinfo {author} {\bibfnamefont
  {T.}~\bibnamefont {Nojima}}, \bibinfo {author} {\bibfnamefont
  {H.}~\bibnamefont {Aoki}}, \bibinfo {author} {\bibfnamefont {Y.}~\bibnamefont
  {Iwasa}}, \ and\ \bibinfo {author} {\bibfnamefont {M.}~\bibnamefont
  {Kawasaki}},\ }\href@noop {} {\bibfield  {journal} {\bibinfo  {journal}
  {Nature materials}\ }\textbf {\bibinfo {volume} {7}},\ \bibinfo {pages} {855}
  (\bibinfo {year} {2008})}\BibitemShut {NoStop}%
\bibitem [{\citenamefont {Mooradian}\ and\ \citenamefont
  {Wright}(1966)}]{Mooradian1966}%
  \BibitemOpen
  \bibfield  {author} {\bibinfo {author} {\bibfnamefont {A.}~\bibnamefont
  {Mooradian}}\ and\ \bibinfo {author} {\bibfnamefont {G.~B.}\ \bibnamefont
  {Wright}},\ }\href {\doibase 10.1103/PhysRevLett.16.999} {\bibfield
  {journal} {\bibinfo  {journal} {Phys. Rev. Lett.}\ }\textbf {\bibinfo
  {volume} {16}},\ \bibinfo {pages} {999} (\bibinfo {year} {1966})}\BibitemShut
  {NoStop}%
\bibitem [{\citenamefont {Cohen}(1969)}]{cohen1969superconductivity}%
  \BibitemOpen
  \bibfield  {author} {\bibinfo {author} {\bibfnamefont {M.~L.}\ \bibnamefont
  {Cohen}},\ }\href@noop {} {\emph {\bibinfo {title} {SUPERCONDUCTIVITY IN
  LOW-CARRIER-DENSITY SYSTEMS: DEGENERATE SEMICONDUCTORS.}}},\ edited by\
  \bibinfo {editor} {\bibfnamefont {R.~D.}\ \bibnamefont {Parks}}\ (\bibinfo
  {publisher} {New York, Marcel Dekker, Inc.},\ \bibinfo {year}
  {1969})\BibitemShut {NoStop}%
\bibitem [{\citenamefont {Mahan}(2013)}]{mahan2013many}%
  \BibitemOpen
  \bibfield  {author} {\bibinfo {author} {\bibfnamefont {G.~D.}\ \bibnamefont
  {Mahan}},\ }\href@noop {} {\emph {\bibinfo {title} {Many-particle physics}}}\
  (\bibinfo  {publisher} {Springer Science \& Business Media},\ \bibinfo {year}
  {2013})\BibitemShut {NoStop}%
\bibitem [{\citenamefont {Appel}(1969)}]{Appel1969}%
  \BibitemOpen
  \bibfield  {author} {\bibinfo {author} {\bibfnamefont {J.}~\bibnamefont
  {Appel}},\ }\href {\doibase 10.1103/PhysRev.180.508} {\bibfield  {journal}
  {\bibinfo  {journal} {Phys. Rev.}\ }\textbf {\bibinfo {volume} {180}},\
  \bibinfo {pages} {508} (\bibinfo {year} {1969})}\BibitemShut {NoStop}%
\bibitem [{\citenamefont {Eagles}(1969)}]{Eagels1969}%
  \BibitemOpen
  \bibfield  {author} {\bibinfo {author} {\bibfnamefont {D.~M.}\ \bibnamefont
  {Eagles}},\ }\href {\doibase 10.1103/PhysRev.186.456} {\bibfield  {journal}
  {\bibinfo  {journal} {Phys. Rev.}\ }\textbf {\bibinfo {volume} {186}},\
  \bibinfo {pages} {456} (\bibinfo {year} {1969})}\BibitemShut {NoStop}%
\bibitem [{\citenamefont {Takada}(1978)}]{Takada1978}%
  \BibitemOpen
  \bibfield  {author} {\bibinfo {author} {\bibfnamefont {Y.}~\bibnamefont
  {Takada}},\ }\href {\doibase 10.1143/JPSJ.45.786} {\bibfield  {journal}
  {\bibinfo  {journal} {Journal of the Physical Society of Japan}\ }\textbf
  {\bibinfo {volume} {45}},\ \bibinfo {pages} {786} (\bibinfo {year}
  {1978})}\BibitemShut {NoStop}%
\bibitem [{\citenamefont {Rietschel}\ and\ \citenamefont
  {Sham}(1983)}]{Rietschel1983}%
  \BibitemOpen
  \bibfield  {author} {\bibinfo {author} {\bibfnamefont {H.}~\bibnamefont
  {Rietschel}}\ and\ \bibinfo {author} {\bibfnamefont {L.~J.}\ \bibnamefont
  {Sham}},\ }\href {\doibase 10.1103/PhysRevB.28.5100} {\bibfield  {journal}
  {\bibinfo  {journal} {Phys. Rev. B}\ }\textbf {\bibinfo {volume} {28}},\
  \bibinfo {pages} {5100} (\bibinfo {year} {1983})}\BibitemShut {NoStop}%
\bibitem [{\citenamefont {Eliashberg}(1960)}]{Eliashberg1960}%
  \BibitemOpen
  \bibfield  {author} {\bibinfo {author} {\bibfnamefont {G.~M.}\ \bibnamefont
  {Eliashberg}},\ }\href@noop {} {\bibfield  {journal} {\bibinfo  {journal}
  {Sov. Phys. Sol. JETP}\ }\textbf {\bibinfo {volume} {11}},\ \bibinfo {pages}
  {696} (\bibinfo {year} {1960})}\BibitemShut {NoStop}%
\bibitem [{\citenamefont {Margine}\ and\ \citenamefont
  {Giustino}(2013)}]{Margine2013}%
  \BibitemOpen
  \bibfield  {author} {\bibinfo {author} {\bibfnamefont {E.~R.}\ \bibnamefont
  {Margine}}\ and\ \bibinfo {author} {\bibfnamefont {F.}~\bibnamefont
  {Giustino}},\ }\href {\doibase 10.1103/PhysRevB.87.024505} {\bibfield
  {journal} {\bibinfo  {journal} {Phys. Rev. B}\ }\textbf {\bibinfo {volume}
  {87}},\ \bibinfo {pages} {024505} (\bibinfo {year} {2013})}\BibitemShut
  {NoStop}%
\bibitem [{\citenamefont {Meevasana}\ \emph {et~al.}(2010)\citenamefont
  {Meevasana}, \citenamefont {Zhou}, \citenamefont {Moritz}, \citenamefont
  {Chen}, \citenamefont {He}, \citenamefont {Fujimori}, \citenamefont {Lu},
  \citenamefont {Mo}, \citenamefont {Moore}, \citenamefont {Baumberger},
  \citenamefont {Devereaux}, \citenamefont {van~der Marel}, \citenamefont
  {Nagaosa}, \citenamefont {Zaanen},\ and\ \citenamefont
  {Shen}}]{Meevasana2010}%
  \BibitemOpen
  \bibfield  {author} {\bibinfo {author} {\bibfnamefont {W.}~\bibnamefont
  {Meevasana}}, \bibinfo {author} {\bibfnamefont {X.~J.}\ \bibnamefont {Zhou}},
  \bibinfo {author} {\bibfnamefont {B.}~\bibnamefont {Moritz}}, \bibinfo
  {author} {\bibfnamefont {C.-C.}\ \bibnamefont {Chen}}, \bibinfo {author}
  {\bibfnamefont {R.~H.}\ \bibnamefont {He}}, \bibinfo {author} {\bibfnamefont
  {S.-I.}\ \bibnamefont {Fujimori}}, \bibinfo {author} {\bibfnamefont {D.~H.}\
  \bibnamefont {Lu}}, \bibinfo {author} {\bibfnamefont {S.-K.}\ \bibnamefont
  {Mo}}, \bibinfo {author} {\bibfnamefont {R.~G.}\ \bibnamefont {Moore}},
  \bibinfo {author} {\bibfnamefont {F.}~\bibnamefont {Baumberger}}, \bibinfo
  {author} {\bibfnamefont {T.~P.}\ \bibnamefont {Devereaux}}, \bibinfo {author}
  {\bibfnamefont {D.}~\bibnamefont {van~der Marel}}, \bibinfo {author}
  {\bibfnamefont {N.}~\bibnamefont {Nagaosa}}, \bibinfo {author} {\bibfnamefont
  {J.}~\bibnamefont {Zaanen}}, \ and\ \bibinfo {author} {\bibfnamefont {Z.-X.}\
  \bibnamefont {Shen}},\ }\href
  {http://stacks.iop.org/1367-2630/12/i=2/a=023004} {\bibfield  {journal}
  {\bibinfo  {journal} {New Journal of Physics}\ }\textbf {\bibinfo {volume}
  {12}},\ \bibinfo {pages} {023004} (\bibinfo {year} {2010})}\BibitemShut
  {NoStop}%
\bibitem [{\citenamefont {Richter}\ \emph {et~al.}(2013)\citenamefont
  {Richter}, \citenamefont {Boschker}, \citenamefont {Dietsche}, \citenamefont
  {Fillis-Tsirakis}, \citenamefont {Jany}, \citenamefont {Loder}, \citenamefont
  {Kourkoutis}, \citenamefont {Muller}, \citenamefont {Kirtley}, \citenamefont
  {Schneider} \emph {et~al.}}]{richter2013interface}%
  \BibitemOpen
  \bibfield  {author} {\bibinfo {author} {\bibfnamefont {C.}~\bibnamefont
  {Richter}}, \bibinfo {author} {\bibfnamefont {H.}~\bibnamefont {Boschker}},
  \bibinfo {author} {\bibfnamefont {W.}~\bibnamefont {Dietsche}}, \bibinfo
  {author} {\bibfnamefont {E.}~\bibnamefont {Fillis-Tsirakis}}, \bibinfo
  {author} {\bibfnamefont {R.}~\bibnamefont {Jany}}, \bibinfo {author}
  {\bibfnamefont {F.}~\bibnamefont {Loder}}, \bibinfo {author} {\bibfnamefont
  {L.}~\bibnamefont {Kourkoutis}}, \bibinfo {author} {\bibfnamefont
  {D.}~\bibnamefont {Muller}}, \bibinfo {author} {\bibfnamefont
  {J.}~\bibnamefont {Kirtley}}, \bibinfo {author} {\bibfnamefont
  {C.}~\bibnamefont {Schneider}},  \emph {et~al.},\ }\href@noop {} {\bibfield
  {journal} {\bibinfo  {journal} {Nature}\ }\textbf {\bibinfo {volume} {502}},\
  \bibinfo {pages} {528} (\bibinfo {year} {2013})}\BibitemShut {NoStop}%
\bibitem [{\citenamefont {Ohtomo}\ and\ \citenamefont
  {Hwang}(2004)}]{ohtomo2004high}%
  \BibitemOpen
  \bibfield  {author} {\bibinfo {author} {\bibfnamefont {A.}~\bibnamefont
  {Ohtomo}}\ and\ \bibinfo {author} {\bibfnamefont {H.}~\bibnamefont {Hwang}},\
  }\href@noop {} {\bibfield  {journal} {\bibinfo  {journal} {Nature}\ }\textbf
  {\bibinfo {volume} {427}},\ \bibinfo {pages} {423} (\bibinfo {year}
  {2004})}\BibitemShut {NoStop}%
\bibitem [{\citenamefont {Santander-Syro}\ \emph {et~al.}(2011)\citenamefont
  {Santander-Syro}, \citenamefont {Copie}, \citenamefont {Kondo}, \citenamefont
  {Fortuna}, \citenamefont {Pailhes}, \citenamefont {Weht}, \citenamefont
  {Qiu}, \citenamefont {Bertran}, \citenamefont {Nicolaou}, \citenamefont
  {Taleb-Ibrahimi} \emph {et~al.}}]{santander2011two}%
  \BibitemOpen
  \bibfield  {author} {\bibinfo {author} {\bibfnamefont {A.}~\bibnamefont
  {Santander-Syro}}, \bibinfo {author} {\bibfnamefont {O.}~\bibnamefont
  {Copie}}, \bibinfo {author} {\bibfnamefont {T.}~\bibnamefont {Kondo}},
  \bibinfo {author} {\bibfnamefont {F.}~\bibnamefont {Fortuna}}, \bibinfo
  {author} {\bibfnamefont {S.}~\bibnamefont {Pailhes}}, \bibinfo {author}
  {\bibfnamefont {R.}~\bibnamefont {Weht}}, \bibinfo {author} {\bibfnamefont
  {X.}~\bibnamefont {Qiu}}, \bibinfo {author} {\bibfnamefont {F.}~\bibnamefont
  {Bertran}}, \bibinfo {author} {\bibfnamefont {A.}~\bibnamefont {Nicolaou}},
  \bibinfo {author} {\bibfnamefont {A.}~\bibnamefont {Taleb-Ibrahimi}},  \emph
  {et~al.},\ }\href@noop {} {\bibfield  {journal} {\bibinfo  {journal}
  {Nature}\ }\textbf {\bibinfo {volume} {469}},\ \bibinfo {pages} {189}
  (\bibinfo {year} {2011})}\BibitemShut {NoStop}%
\bibitem [{\citenamefont {Worlock}\ and\ \citenamefont
  {Fleury}(1967)}]{worlock1967electric}%
  \BibitemOpen
  \bibfield  {author} {\bibinfo {author} {\bibfnamefont {J.}~\bibnamefont
  {Worlock}}\ and\ \bibinfo {author} {\bibfnamefont {P.}~\bibnamefont
  {Fleury}},\ }\href@noop {} {\bibfield  {journal} {\bibinfo  {journal}
  {Physical Review Letters}\ }\textbf {\bibinfo {volume} {19}},\ \bibinfo
  {pages} {1176} (\bibinfo {year} {1967})}\BibitemShut {NoStop}%
\bibitem [{\citenamefont {Sirenko}\ \emph {et~al.}(2000)\citenamefont
  {Sirenko}, \citenamefont {Bernhard}, \citenamefont {Golnik}, \citenamefont
  {Clark}, \citenamefont {Hao}, \citenamefont {Si},\ and\ \citenamefont
  {Xi}}]{sirenko2000soft}%
  \BibitemOpen
  \bibfield  {author} {\bibinfo {author} {\bibfnamefont {A.}~\bibnamefont
  {Sirenko}}, \bibinfo {author} {\bibfnamefont {C.}~\bibnamefont {Bernhard}},
  \bibinfo {author} {\bibfnamefont {A.}~\bibnamefont {Golnik}}, \bibinfo
  {author} {\bibfnamefont {A.~M.}\ \bibnamefont {Clark}}, \bibinfo {author}
  {\bibfnamefont {J.}~\bibnamefont {Hao}}, \bibinfo {author} {\bibfnamefont
  {W.}~\bibnamefont {Si}}, \ and\ \bibinfo {author} {\bibfnamefont
  {X.}~\bibnamefont {Xi}},\ }\href@noop {} {\bibfield  {journal} {\bibinfo
  {journal} {Nature}\ }\textbf {\bibinfo {volume} {404}},\ \bibinfo {pages}
  {373} (\bibinfo {year} {2000})}\BibitemShut {NoStop}%
\bibitem [{\citenamefont {Bell}\ \emph {et~al.}(2009)\citenamefont {Bell},
  \citenamefont {Harashima}, \citenamefont {Kozuka}, \citenamefont {Kim},
  \citenamefont {Kim}, \citenamefont {Hikita},\ and\ \citenamefont
  {Hwang}}]{Bell2009}%
  \BibitemOpen
  \bibfield  {author} {\bibinfo {author} {\bibfnamefont {C.}~\bibnamefont
  {Bell}}, \bibinfo {author} {\bibfnamefont {S.}~\bibnamefont {Harashima}},
  \bibinfo {author} {\bibfnamefont {Y.}~\bibnamefont {Kozuka}}, \bibinfo
  {author} {\bibfnamefont {M.}~\bibnamefont {Kim}}, \bibinfo {author}
  {\bibfnamefont {B.~G.}\ \bibnamefont {Kim}}, \bibinfo {author} {\bibfnamefont
  {Y.}~\bibnamefont {Hikita}}, \ and\ \bibinfo {author} {\bibfnamefont {H.~Y.}\
  \bibnamefont {Hwang}},\ }\href {\doibase 10.1103/PhysRevLett.103.226802}
  {\bibfield  {journal} {\bibinfo  {journal} {Phys. Rev. Lett.}\ }\textbf
  {\bibinfo {volume} {103}},\ \bibinfo {pages} {226802} (\bibinfo {year}
  {2009})}\BibitemShut {NoStop}%
\bibitem [{\citenamefont {Schrieffer}\ \emph {et~al.}(1963)\citenamefont
  {Schrieffer}, \citenamefont {Scalapino},\ and\ \citenamefont
  {Wilkins}}]{Schrieffer1963}%
  \BibitemOpen
  \bibfield  {author} {\bibinfo {author} {\bibfnamefont {J.~R.}\ \bibnamefont
  {Schrieffer}}, \bibinfo {author} {\bibfnamefont {D.~J.}\ \bibnamefont
  {Scalapino}}, \ and\ \bibinfo {author} {\bibfnamefont {J.~W.}\ \bibnamefont
  {Wilkins}},\ }\href {\doibase 10.1103/PhysRevLett.10.336} {\bibfield
  {journal} {\bibinfo  {journal} {Phys. Rev. Lett.}\ }\textbf {\bibinfo
  {volume} {10}},\ \bibinfo {pages} {336} (\bibinfo {year} {1963})}\BibitemShut
  {NoStop}%
\bibitem [{\citenamefont {Beach}\ \emph {et~al.}(2000)\citenamefont {Beach},
  \citenamefont {Gooding},\ and\ \citenamefont {Marsiglio}}]{Beach2000}%
  \BibitemOpen
  \bibfield  {author} {\bibinfo {author} {\bibfnamefont {K.}~\bibnamefont
  {Beach}}, \bibinfo {author} {\bibfnamefont {R.}~\bibnamefont {Gooding}}, \
  and\ \bibinfo {author} {\bibfnamefont {F.}~\bibnamefont {Marsiglio}},\ }\href
  {\doibase 10.1103/PhysRevB.61.5147} {\bibfield  {journal} {\bibinfo
  {journal} {Physical Review B}\ }\textbf {\bibinfo {volume} {61}},\ \bibinfo
  {pages} {5147} (\bibinfo {year} {2000})},\ \Eprint
  {http://arxiv.org/abs/9908477} {arXiv:9908477 [cond-mat]} \BibitemShut
  {NoStop}%
\bibitem [{\citenamefont {Venturini}\ \emph {et~al.}(2003)\citenamefont
  {Venturini}, \citenamefont {Samara},\ and\ \citenamefont
  {Kleemann}}]{Venturini2003}%
  \BibitemOpen
  \bibfield  {author} {\bibinfo {author} {\bibfnamefont {E.~L.}\ \bibnamefont
  {Venturini}}, \bibinfo {author} {\bibfnamefont {G.~A.}\ \bibnamefont
  {Samara}}, \ and\ \bibinfo {author} {\bibfnamefont {W.}~\bibnamefont
  {Kleemann}},\ }\href {\doibase 10.1103/PhysRevB.67.214102} {\bibfield
  {journal} {\bibinfo  {journal} {Phys. Rev. B}\ }\textbf {\bibinfo {volume}
  {67}},\ \bibinfo {pages} {214102} (\bibinfo {year} {2003})}\BibitemShut
  {NoStop}%
\bibitem [{\citenamefont {de~Lima}\ \emph {et~al.}(2015)\citenamefont
  {de~Lima}, \citenamefont {da~Luz}, \citenamefont {Oliveira}, \citenamefont
  {Alves}, \citenamefont {dos Santos}, \citenamefont {Jomard}, \citenamefont
  {Sidis}, \citenamefont {Bourges}, \citenamefont {Harms}, \citenamefont
  {Grams} \emph {et~al.}}]{de2015interplay}%
  \BibitemOpen
  \bibfield  {author} {\bibinfo {author} {\bibfnamefont {B.}~\bibnamefont
  {de~Lima}}, \bibinfo {author} {\bibfnamefont {M.}~\bibnamefont {da~Luz}},
  \bibinfo {author} {\bibfnamefont {F.}~\bibnamefont {Oliveira}}, \bibinfo
  {author} {\bibfnamefont {L.}~\bibnamefont {Alves}}, \bibinfo {author}
  {\bibfnamefont {C.}~\bibnamefont {dos Santos}}, \bibinfo {author}
  {\bibfnamefont {F.}~\bibnamefont {Jomard}}, \bibinfo {author} {\bibfnamefont
  {Y.}~\bibnamefont {Sidis}}, \bibinfo {author} {\bibfnamefont
  {P.}~\bibnamefont {Bourges}}, \bibinfo {author} {\bibfnamefont
  {S.}~\bibnamefont {Harms}}, \bibinfo {author} {\bibfnamefont
  {C.}~\bibnamefont {Grams}},  \emph {et~al.},\ }\href@noop {} {\bibfield
  {journal} {\bibinfo  {journal} {Physical Review B}\ }\textbf {\bibinfo
  {volume} {91}},\ \bibinfo {pages} {045108} (\bibinfo {year}
  {2015})}\BibitemShut {NoStop}%
\bibitem [{\citenamefont {Eagles}(1965)}]{eagles1965polaron}%
  \BibitemOpen
  \bibfield  {author} {\bibinfo {author} {\bibfnamefont {D.}~\bibnamefont
  {Eagles}},\ }\href@noop {} {\bibfield  {journal} {\bibinfo  {journal}
  {Journal of Physics and Chemistry of Solids}\ }\textbf {\bibinfo {volume}
  {26}},\ \bibinfo {pages} {672} (\bibinfo {year} {1965})}\BibitemShut
  {NoStop}%
\bibitem [{\citenamefont {Hor}\ \emph {et~al.}(2010)\citenamefont {Hor},
  \citenamefont {Williams}, \citenamefont {Checkelsky}, \citenamefont
  {Roushan}, \citenamefont {Seo}, \citenamefont {Xu}, \citenamefont
  {Zandbergen}, \citenamefont {Yazdani}, \citenamefont {Ong},\ and\
  \citenamefont {Cava}}]{Hor2010}%
  \BibitemOpen
  \bibfield  {author} {\bibinfo {author} {\bibfnamefont {Y.~S.}\ \bibnamefont
  {Hor}}, \bibinfo {author} {\bibfnamefont {A.~J.}\ \bibnamefont {Williams}},
  \bibinfo {author} {\bibfnamefont {J.~G.}\ \bibnamefont {Checkelsky}},
  \bibinfo {author} {\bibfnamefont {P.}~\bibnamefont {Roushan}}, \bibinfo
  {author} {\bibfnamefont {J.}~\bibnamefont {Seo}}, \bibinfo {author}
  {\bibfnamefont {Q.}~\bibnamefont {Xu}}, \bibinfo {author} {\bibfnamefont
  {H.~W.}\ \bibnamefont {Zandbergen}}, \bibinfo {author} {\bibfnamefont
  {A.}~\bibnamefont {Yazdani}}, \bibinfo {author} {\bibfnamefont {N.~P.}\
  \bibnamefont {Ong}}, \ and\ \bibinfo {author} {\bibfnamefont {R.~J.}\
  \bibnamefont {Cava}},\ }\href {\doibase 10.1103/PhysRevLett.104.057001}
  {\bibfield  {journal} {\bibinfo  {journal} {Phys. Rev. Lett.}\ }\textbf
  {\bibinfo {volume} {104}},\ \bibinfo {pages} {057001} (\bibinfo {year}
  {2010})}\BibitemShut {NoStop}%
\bibitem [{\citenamefont {Wray}\ \emph {et~al.}(2010)\citenamefont {Wray},
  \citenamefont {Xu}, \citenamefont {Xia}, \citenamefont {San~Hor},
  \citenamefont {Qian}, \citenamefont {Fedorov}, \citenamefont {Lin},
  \citenamefont {Bansil}, \citenamefont {Cava},\ and\ \citenamefont
  {Hasan}}]{wray2010observation}%
  \BibitemOpen
  \bibfield  {author} {\bibinfo {author} {\bibfnamefont {L.~A.}\ \bibnamefont
  {Wray}}, \bibinfo {author} {\bibfnamefont {S.-Y.}\ \bibnamefont {Xu}},
  \bibinfo {author} {\bibfnamefont {Y.}~\bibnamefont {Xia}}, \bibinfo {author}
  {\bibfnamefont {Y.}~\bibnamefont {San~Hor}}, \bibinfo {author} {\bibfnamefont
  {D.}~\bibnamefont {Qian}}, \bibinfo {author} {\bibfnamefont {A.~V.}\
  \bibnamefont {Fedorov}}, \bibinfo {author} {\bibfnamefont {H.}~\bibnamefont
  {Lin}}, \bibinfo {author} {\bibfnamefont {A.}~\bibnamefont {Bansil}},
  \bibinfo {author} {\bibfnamefont {R.~J.}\ \bibnamefont {Cava}}, \ and\
  \bibinfo {author} {\bibfnamefont {M.~Z.}\ \bibnamefont {Hasan}},\ }\href@noop
  {} {\bibfield  {journal} {\bibinfo  {journal} {Nature Physics}\ }\textbf
  {\bibinfo {volume} {6}},\ \bibinfo {pages} {855} (\bibinfo {year}
  {2010})}\BibitemShut {NoStop}%
\bibitem [{\citenamefont {Kriener}\ \emph {et~al.}(2011)\citenamefont
  {Kriener}, \citenamefont {Segawa}, \citenamefont {Ren}, \citenamefont
  {Sasaki},\ and\ \citenamefont {Ando}}]{kriener2011bulk}%
  \BibitemOpen
  \bibfield  {author} {\bibinfo {author} {\bibfnamefont {M.}~\bibnamefont
  {Kriener}}, \bibinfo {author} {\bibfnamefont {K.}~\bibnamefont {Segawa}},
  \bibinfo {author} {\bibfnamefont {Z.}~\bibnamefont {Ren}}, \bibinfo {author}
  {\bibfnamefont {S.}~\bibnamefont {Sasaki}}, \ and\ \bibinfo {author}
  {\bibfnamefont {Y.}~\bibnamefont {Ando}},\ }\href@noop {} {\bibfield
  {journal} {\bibinfo  {journal} {Physical review letters}\ }\textbf {\bibinfo
  {volume} {106}},\ \bibinfo {pages} {127004} (\bibinfo {year}
  {2011})}\BibitemShut {NoStop}%
\bibitem [{\citenamefont {Liu}\ \emph {et~al.}(2015)\citenamefont {Liu},
  \citenamefont {Yao}, \citenamefont {Shao}, \citenamefont {Zuo}, \citenamefont
  {Pi}, \citenamefont {Tan}, \citenamefont {Zhang},\ and\ \citenamefont
  {Zhang}}]{liu2015turning}%
  \BibitemOpen
  \bibfield  {author} {\bibinfo {author} {\bibfnamefont {Z.}~\bibnamefont
  {Liu}}, \bibinfo {author} {\bibfnamefont {X.}~\bibnamefont {Yao}}, \bibinfo
  {author} {\bibfnamefont {J.}~\bibnamefont {Shao}}, \bibinfo {author}
  {\bibfnamefont {M.}~\bibnamefont {Zuo}}, \bibinfo {author} {\bibfnamefont
  {L.}~\bibnamefont {Pi}}, \bibinfo {author} {\bibfnamefont {S.}~\bibnamefont
  {Tan}}, \bibinfo {author} {\bibfnamefont {C.}~\bibnamefont {Zhang}}, \ and\
  \bibinfo {author} {\bibfnamefont {Y.}~\bibnamefont {Zhang}},\ }\href@noop {}
  {\bibfield  {journal} {\bibinfo  {journal} {arXiv preprint arXiv:1502.01105}\
  } (\bibinfo {year} {2015})}\BibitemShut {NoStop}%
\bibitem [{\citenamefont {Zhong}\ \emph {et~al.}(1994)\citenamefont {Zhong},
  \citenamefont {King-Smith},\ and\ \citenamefont {Vanderbilt}}]{Zhong1994}%
  \BibitemOpen
  \bibfield  {author} {\bibinfo {author} {\bibfnamefont {W.}~\bibnamefont
  {Zhong}}, \bibinfo {author} {\bibfnamefont {R.~D.}\ \bibnamefont
  {King-Smith}}, \ and\ \bibinfo {author} {\bibfnamefont {D.}~\bibnamefont
  {Vanderbilt}},\ }\href {\doibase 10.1103/PhysRevLett.72.3618} {\bibfield
  {journal} {\bibinfo  {journal} {Phys. Rev. Lett.}\ }\textbf {\bibinfo
  {volume} {72}},\ \bibinfo {pages} {3618} (\bibinfo {year}
  {1994})}\BibitemShut {NoStop}%
\bibitem [{\citenamefont {Kozii}\ and\ \citenamefont {Fu}(2015)}]{Kozii2015}%
  \BibitemOpen
  \bibfield  {author} {\bibinfo {author} {\bibfnamefont {V.}~\bibnamefont
  {Kozii}}\ and\ \bibinfo {author} {\bibfnamefont {L.}~\bibnamefont {Fu}},\
  }\href {\doibase 10.1103/PhysRevLett.115.207002} {\bibfield  {journal}
  {\bibinfo  {journal} {Phys. Rev. Lett.}\ }\textbf {\bibinfo {volume} {115}},\
  \bibinfo {pages} {207002} (\bibinfo {year} {2015})}\BibitemShut {NoStop}%
\end{thebibliography}
\end{document}